\documentclass[article]{jfm}
\usepackage{setspace}
\usepackage{graphicx}
\usepackage{newtxtext}
\usepackage{newtxmath}
\usepackage{natbib}
\usepackage{arydshln}
\usepackage{hyperref}
\hypersetup{
    colorlinks = true,
    urlcolor   = blue,
    citecolor  = black,
}
\newcommand{\RomanNumeralCaps}[1]

\title{Magneto-Stokes Flow in a Shallow Free-Surface Annulus}

\author{Cy S. David\aff{1}
  \corresp{\email{cysdavid@ucla.edu}},
  Eric W. Hester\aff{2},
  Yufan Xu\aff{1,3}
 \and Jonathan M. Aurnou\aff{1}}

\affiliation{\aff{1}Department of Earth, Planetary, and Space Sciences, University of California, Los Angeles, CA 90095, USA
\aff{2}Department of Mathematics, University of California, Los Angeles, CA 90095, USA
\aff{3}Princeton Plasma Physics Laboratory, Princeton, NJ 08540, USA}

\begin{document}
\maketitle

\begin{abstract} 
In this study, we analyse ``magneto-Stokes'' flow, a fundamental magnetohydrodynamic (MHD) flow that shares the cylindrical-annular geometry of the Taylor-Couette cell, but uses applied electromagnetic forces to circulate a free-surface layer of electrolyte at low Reynolds numbers. The first complete, analytical solution for time-dependent magneto-Stokes flow is presented and validated with coupled laboratory and numerical experiments. Three regimes are distinguished (shallow-layer, transitional, and deep-layer flow regimes), and their influence on the efficiency of microscale mixing is clarified. The solution in the shallow-layer limit belongs to a newly-identified class of MHD potential flows, and thus induces mixing without the aid of axial vorticity. We show that these shallow-layer magneto-Stokes flows can still augment mixing in distinct Taylor dispersion and advection-dominated mixing regimes. The existence of enhanced mixing across all three distinguished flow regimes is predicted by asymptotic scaling laws and supported by three-dimensional numerical simulations. Mixing enhancement is initiated with the least electromagnetic forcing in channels with order-unity depth-to-gap-width ratios. If the strength of the electromagnetic forcing is not a constraint, then shallow-layer flows can still yield the shortest mixing times in the advection-dominated limit. Our robust description of momentum evolution and mixing of passive tracers makes the annular magneto-Stokes system fit for use as an MHD reference flow.
\end{abstract}


\section{Introduction}\label{sec:intro}
The year 2023 marks the 100th anniversary of G.I. Taylor's seminal publication \citep{taylor_viii_1923} on flow confined between two rotating concentric cylinders (\textit{Taylor-Couette flow}) and the 70th anniversary of his work \citep{taylor_dispersion_1953} on enhanced mixing in shear flows (\textit{Taylor dispersion}). With these two landmark papers in mind, we develop a magnetohydrodynamic (MHD) reference flow inspired by the Taylor-Couette system that enhances mixing at low Reynolds numbers via Taylor dispersion.

Our MHD modification of the Taylor-Couette cell uses electromagnetic body forces rather than viscous traction to drive motion; the usually rotating sidewalls of the cylindrical annulus are fixed and made electrically-conducting. The base is kept electrically-insulating, while the lid is removed to allow free-surface flow. An applied axial magnetic field and radial electric current drive azimuthal flow of an electrolyte, for which (1) magnetic induction is small compared to magnetic diffusion (low \textit{magnetic Reynolds number}, $\textit{Rm}$), (2) magnetic drag is small relative to viscous forces (low \textit{Hartmann number}, $\textit{Ha}$), and (3) inertia is small compared to viscous forces (low \textit{Reynolds number}, $\Rey$). Under these conditions, the azimuthal momentum balance is dominated by the Lorentz force and viscous drag due to the channel sidewalls and base. We term the resulting circulatory motion ``annular magneto-Stokes flow.''

Similar MHD flows through cylindrical annular ducts with conducting sidewalls and axial magnetic field have been considered since the works of \cite{early_supersonic_1950}, \cite{ anderson_hydromagnetic_1959}, and \cite{braginsky_magnetohydrodynamics_1959}. Early modelling efforts focus on the high-Hartmann number limit for analytical convenience \citep{hunt_magnetohydrodynamic_1965} and relevance to liquid metal flows \citep{baylis_mhd_1971}. Later numerical and laboratory works have surveyed a broader range of Hartmann numbers in closed annular ducts \citep{poye_scaling_2020,vernet_turbulence_2021}, analysed the stability of Hartmann layers \citep{moresco_experimental_2004}, and studied the effects of modified electric boundary conditions \citep{stelzer_experimental_2015-1,stelzer_experimental_2015} in cylindrical-annular MHD flows. Recently, liquid metal experiments \citep{vernet_angular_2022} in a similar geometry have accessed a regime of Keplerian turbulence representative of flows in astrophysical disks.

In contrast to the above studies involving large-scale liquid metal systems, applications of MHD to microfluidic mixing devices have generated interest in low-Hartmann number flows \citep{west_structuring_2003,khalzov_calculation_2006} appropriate for electrolytes. Initial efforts to model annular magneto-Stokes flow \citep{gleeson_magnetohydrodynamic_2002,gleeson_modelling_2004,digilov_making_2007} assumed an infinitely-deep layer, arriving at a two-dimensional (2D) asymptotic solution. \cite{perez-barrera_instability_2015,klapp_analysis_2016} later considered channels of finite depth, solving specifically for the vertically-averaged velocity profile $\langle u_\theta\rangle_z (r)$. Following this, \cite{ortiz-perez_magnetohydrodynamic_2017} and \cite{valenzuela-delgado_electrolyte_2018,valenzuela-delgado_theoretical_2018} used a (semi-analytical) Galerkin approximation to predict steady, axisymmetric flow over radius and depth, $u_\theta(r,z)$. A fully analytical solution $u_\theta(r,z,t)$ for time-dependent annular magneto-Stokes flow does not exist in the literature, to our knowledge, despite the simplicity of the governing equation. Further, there is little discussion of the range of channel geometries for which the deep-layer approximation is valid. Yet, the assumption of infinite depth has been made for engineering problems involving shallow-layer flows (e.g., 
\citealt{west_structuring_2003}) with strong vertical shear that depart greatly from the 2D deep-layer solution.

In this study, we unify and extend these previous efforts by: (i) providing the first, complete analytical solution for time-dependent flow $u_\theta(r,z,t)$ in a channel of arbitrary depth (\S\ref{sec:theory}), which we validate with laboratory experiments and direct numerical simulations (DNS) (\S\ref{sec:methods}, \ref{sec:results}); (ii) correctly distinguishing deep, transitional, and shallow-layer flow regimes in terms of the appropriate geometric parameter (\S\ref{sec:theory}); (iii) deriving the shallow-layer asymptotic solution (\S\ref{sec:theory}); (iv) applying these findings to the design of a microfluidic mixer (\S\ref{sec:apps}); and (v) showing that the onset of shear-enhanced mixing occurs with the least electromagnetic forcing in the transitional flow regime (\S\ref{sec:apps}).
\section{Theory}\label{sec:theory}

\begin{figure}
  \centerline{\includegraphics{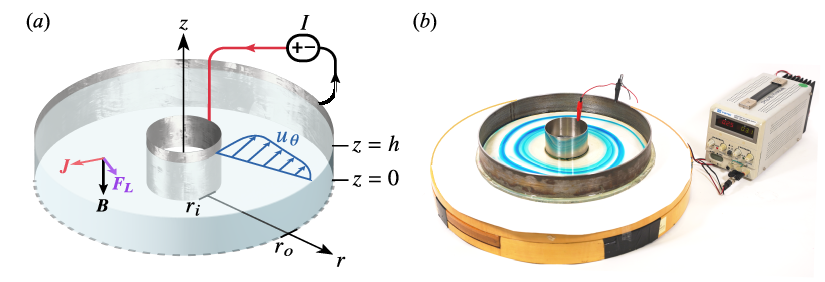}}
  \caption{(\textit{a}) Diagram of the magneto-Stokes system. A power supply controls the electric current $I$ through the fluid layer. Radially outwards ($+\boldsymbol{e}_r$) current density $\boldsymbol{J}$ and downwards ($-\boldsymbol{e}_z$) magnetic field $\boldsymbol{B}$ produce an azimuthal ($+\boldsymbol{e}_\theta$) Lorentz force $\boldsymbol{F}$ on the fluid. (\textit{b}) Photograph of the channel used in laboratory experiments, with flow visualised by blue dye. The channel rests atop a wooden case of permanent magnets, which is replaced by a solenoid electromagnet (not pictured) for Cases I-IV discussed in \S\ref{sec:methods}. }
\label{fig:device}
\end{figure}

\subsection {Axisymmetric governing equations }\label{sec:theory:gov}
We consider a free-surface layer of conducting fluid of depth $h$ in the annular gap between two cylindrical electrodes of radius $r_i$ and $r_o$ ($r_i<r_o$). A controlled current $I$ runs through the fluid from inner to outer electrode, and the entire annulus is subject to a vertical, imposed magnetic field $\boldsymbol{B} =  -B_0 \boldsymbol{e_z}$. Figure \ref{fig:device}(\textit{a}) shows a schematic of the annular channel with the imposed magnetic field and current. For a low-conductivity fluid like saltwater and small $B_0$, the magnetic field is quasi-static (e.g., \citealt{davidson_introduction_2016,knaepen_magnetohydrodynamic_2008,favier_quasi-static_2011,verma_anisotropy_2017}) and the total Lorentz force on the fluid may be expressed as the sum of the applied driving force and magnetic drag,
\begin{equation}\label{eqn:LorentzDef}
    \boldsymbol{F}=\boldsymbol{J}\times\boldsymbol{B}=\sigma(\boldsymbol{E}\times\boldsymbol{B}) - \sigma B_0^2(u_r \boldsymbol{e}_r + u_\theta \boldsymbol{e}_\theta),
\end{equation}
after using Ohm's Law to express the current density $\boldsymbol{J}$ in terms of the electrical conductivity of the fluid $\sigma$, the imposed electric field $\boldsymbol{E}$, and the fluid velocity $\boldsymbol{u}$.

Letting $U$ denote a characteristic velocity scale, the magnitude of the magnetic drag ($\sim\sigma B_0^2U$) may be compared to that of the viscous drag ($\sim\varrho\nu U/h^2$) by means of the Hartmann number:
\begin{equation}\label{eqn:Hadef}
    \textit{Ha} = \sqrt{\frac{\sigma B_0^2U}{\varrho\nu U/h^2}} = B_0 h\sqrt{\frac{\sigma}{\varrho\nu}},
\end{equation}
where $\varrho$ and $\nu$ are the density and kinematic viscosity of the fluid, respectively.

In our experiments, $\textit{Ha} \lesssim 10^{-2}$ and thus the only component of current density $\boldsymbol{J}$ that makes a significant contribution to the Lorentz force is $\sigma \boldsymbol{E}$, which may be determined purely from the electric boundary conditions. A DC power supply can control the voltage across the sidewalls such that the total current $I$ is set to a desired value at each time $t$. In this case, current rather than voltage is used as a control parameter, and the Lorentz force is appropriately expressed as
\begin{equation}\label{eqn:LorentzMagStk}
    \boldsymbol{F} = \sigma(\boldsymbol{E}\times\boldsymbol{B})= \frac{B_0 I(t)}{2 \pi r h}\boldsymbol{e_\theta},
\end{equation}
neglecting any fringing of electric field lines due to the finite fluid depth (i.e., assuming that $\partial_z \boldsymbol{E} = 0$).

The resulting circulatory flow is governed by the incompressible equations of motion:
\begin{equation}\label{eqn:MomVec}
    \partial_t \boldsymbol{u} + \boldsymbol{u}\bcdot\bnabla\boldsymbol{u} = -\frac{1}{\varrho}\bnabla p + \nu \nabla^2 \boldsymbol{u}+ \frac{1}{\varrho}\boldsymbol{F},\quad \bnabla\bcdot\boldsymbol{u} = 0,  
\end{equation}
where $p$ is the deviation in pressure from the static pressure field $\varrho g (h-z)$.

\begin{table}
\begin{center}
\def~{\hphantom{0}}
\addtolength{\tabcolsep}{10pt}  
\begin{tabular}{ll}
Symbol                         & Description                                                \\ \hline
$r_i, r_o$                     & Inner and outer sidewall radii [m]                             \\
$h$                            & Depth of fluid [m]                                             \\
$g$                            & Gravitational acceleration [m/s$^2$]                                 \\
$\nu$                          & Kinematic viscosity [m$^2$/s]                                        \\
$\varrho$                         & Density [kg/m$^3$]                                                    \\
$\sigma$                       & Electrical conductivity [S/m]                                    \\
$\gamma$                       & Surface tension [N/m]                                    \\
$\kappa_c$                       & Tracer diffusivity [m$^2$/s]  \\
$\mu_0$                        & Magnetic permeability of free space [kg m s$^{-2}$ A$^{-2}$]                        \\
$I_0$                          & Maximum electric current amplitude [A]                         \\
$B_0$                          & Magnetic flux density (magnitude) [T]                         \\
$\delta_{i}, \delta_{o}, \delta_{b}$ & \begin{tabular}[c]{@{}l@{}}95\% thickness of inner, outer, and basal boundary layers [m], \\ defined in  (\ref{eqn:BLscaling}, \ref{eqn:BLbasalscaling})\end{tabular}                                  \\
\end{tabular}
\caption{Dimensional parameter definitions. Values are given in \S\ref{sec:methods}.}
\label{tab:DimParamDefs}
\end{center}
\end{table}

We scale radial distances $r$ by the outer sidewall radius $r_o$, vertical distances $z$ by the fluid depth $h$, and electric current $I$ by its maximum value $I_0$, defining the nondimensional quantities $\rho = r/r_o$, $\zeta=z/h$, and $\Upsilon = I/I_0$. A velocity scale $U$ may be found by balancing the basal viscous drag ($\sim 2{\nu U}/{h^2}$) and Lorentz forces ($\sim B_0 I_0/[\pi  h \varrho(r_i+r_o)]$) at the surface mid-gap, $z=h$, $r=(r_i+r_o)/2$:
\begin{equation}\label{eq:UMS}
    U = \frac{B_0 I_0 h}{2 \pi  \nu  \varrho (r_i+r_o)}.
\end{equation}
This scale is used to produce the nondimensional velocity $\boldsymbol{\upsilon} = \upsilon_\rho\boldsymbol{e_r} + \upsilon_\theta\boldsymbol{e_\theta} + (h/r_o)\upsilon_\zeta\boldsymbol{e_z}$, with components $\upsilon_\rho = u_r/U$, $\upsilon_\theta = u_\theta/U$, and $\upsilon_\zeta = (r_o/h) u_z/U$. An inertial scale $\varrho {U}^2$ is used to nondimensionalise reduced pressure as $\Pi = p/(\varrho {U}^2)$. Time is nondimensionalised as $\tau = t/T$ by a surface mid-gap advective timescale:
\begin{equation}\label{eqn:Tdef}
    T=\frac{r_i+r_o}{U}.
\end{equation}
In sum, our nondimensionalisation makes the mapping
\begin{equation}
    (\boldsymbol{u}, p, I)(r,\theta,z,t) \mapsto (\boldsymbol{\upsilon}, \Pi, \Upsilon)(\rho,\theta,\zeta,\tau).
\end{equation}

We introduce an \textit{a priori} control Reynolds number $\Rey$ and a magnetic Reynolds number $\textit{Rm}$:
\begin{equation}\label{eqn:Redef}
    \Rey = \frac{h^2/\nu}{T} = \frac{B_0 I_0 h^3}{2 \pi  \nu ^2 \varrho (r_i+r_o)^2} \quad \text{and} \quad \textit{Rm} = \frac{h^2/\eta}{T} = \frac{\nu}{\eta}\Rey,
\end{equation}
where $\eta = (\mu_0 \sigma)^{-1}$ is the fluid's magnetic diffusivity and $\mu_0$ is the magnetic permeability of free space. For $\textit{Rm} \ll 1$, magnetic diffusion dominates over advection, and the quasi-static description of the Lorentz force (\ref{eqn:LorentzMagStk}) is valid (e.g., \citealt{davidson_introduction_2016,knaepen_magnetohydrodynamic_2008,favier_quasi-static_2011,verma_anisotropy_2017}). As defined, $\textit{Rm} \lesssim 10^{-10}$ for our experiments.

Also relevant are the radius ratio $\mathcal{R}$, aspect ratio $\mathcal{H}$, and depth-to-gap-width ratio $\Gamma$:
\begin{equation}\label{eqn:RHdef}
    \mathcal{R} = r_i/r_o, \quad \mathcal{H} = h/r_o, \quad \text{and} \quad \Gamma = h/(r_o-r_i) = \mathcal{H}/(1-\mathcal{R}).
\end{equation}

Symbols for all dimensional parameters are listed in table \ref{tab:DimParamDefs}. Definitions of scales and nondimensional parameters are collected in table \ref{tab:NdimParamDefs}. The full nondimensional axisymmetric equations in scaled cylindrical coordinates ($\rho$,$\theta$,$\zeta$) may be found in Appendix \ref{appA}.

\begin{table}
\begin{center}
\def~{\hphantom{0}}
\addtolength{\tabcolsep}{1.2pt}  
\renewcommand{\arraystretch}{1.4}
\begin{tabular}{llcl}
Symbol                    & Definition                                                        & Eqn.                   & Description                                \\ \hline 
$U$             & ${B_0 h I_0}/[{2 \pi  \nu  \varrho (r_i+r_o)}]$                      & (\ref{eq:UMS})         & Shallow-layer velocity scale [m/s]         \\
$U_\text{deep}$ & ${B_0 I_0 (r_o-r_i)^2}/[{8 \pi  h \nu  \varrho (r_i+r_o)}]$          & (\ref{eqn:Udeepdef})   & Deep-layer velocity scale [m/s]            \\
$u_{\theta,\text{mid-gap}}$ & $U \bar{\upsilon}_\theta(\rho=(1+\mathcal{R})/2,\zeta=1)$          & (\ref{eqn:umidgapdef})   & \renewcommand{\arraystretch}{0.7}\begin{tabular}[c]{@{}l@{}}Surface mid-gap value\\ of steady solution (\ref{eqn:steadySoln}) [m/s]\end{tabular}            \\
$T$                       & $(r_i+r_o)/U$                                           & (\ref{eqn:Tdef})       & Advective timescale [s]                    \\
$T_\text{su}$          & ${4 h^2}/({\pi^2 \nu })$                                          & (\ref{eqn:tausudef})   & Spin-up timescale [s]                      \\
$T_\text{circ}$        & ${\pi(r_i+r_o)}/{u_{\theta,\text{mid-gap}}}$                                    & (\ref{eqn:taucircdef}) & \renewcommand{\arraystretch}{0.7}\begin{tabular}[c]{@{}l@{}}Surface mid-gap\\ circulation time [s]\end{tabular}         \\[5pt] \hdashline[2pt/2pt]\\[-11pt]
$\textit{Ha}$             & $B_0 h \sqrt{\sigma_0/(\varrho \nu)}$                                & (\ref{eqn:Hadef})      & Hartmann number                            \\
$\Rey$                    & $h^2/(\nu T) = {B_0 I_0 h^3}/[{2 \pi  \varrho\nu ^2  (r_i+r_o)^2}]$  & (\ref{eqn:Redef})      & Control Reynolds number                    \\
$\textit{Rm}$             & $\nu\mu_0\sigma_0\Rey $                                           & (\ref{eqn:Redef})      & Magnetic Reynolds number                   \\
$\textit{Fr}$             & $U/\sqrt{g h}$                                          & (\ref{eqn:Frdef})      & Froude number                              \\
$\textit{Bo}$      & $\varrho g h^2/\gamma$                           & (\ref{eqn:Bodef})      & Bond number                              \\
$\mathcal{R}$             & $r_i/r_o$                                                         & (\ref{eqn:RHdef})      & Sidewall radius ratio                      \\
$\mathcal{H}$             & $h/r_o$                                                           & (\ref{eqn:RHdef})      & Aspect ratio                   \\
$\Gamma$             & $h/(r_o-r_i)$                                                           & (\ref{eqn:RHdef})      & Depth-to-gap-width ratio                   \\
$\Delta_i$,$\Delta_o$,$\Delta_b$                & $\delta_{i}/(r_o - r_i)$, $\delta_{o}/(r_o - r_i)$, $\delta_{b}/h$                                       & (\ref{eqn:BLscaling}, \ref{eqn:BLbasalscaling})   & \renewcommand{\arraystretch}{0.7}\begin{tabular}[c]{@{}l@{}}Inner, outer, and basal\\ boundary layer thicknesses
\end{tabular}\\
$\textit{Sc}$     & $\nu/\kappa_c$          & (\ref{eqn:PeDef})       &Schmidt number\\
$\textit{Pe}$     & $\textit{Sc}\Rey=h^2/(\kappa_c T)$          & (\ref{eqn:PeDef})       &Péclet number\\
$\textit{Pe}_{\text{circ}}$     & ${u_{\theta,\text{mid-gap}}r_o^2}/({U \pi h^2}) \; \textit{Pe} \;=\; r_o^2/(\kappa_c T_{\text{circ}})$          & (\ref{eqn:PeCircDef})       &Circulation Péclet number \\                 
$\widetilde{\textit{Pe}}$     & $({2 \pi (1+\mathcal{R})^2}/{\mathcal{H}^3})\;\textit{Pe} \;=\; {B_0 I_0 r_o}/({\varrho \nu \kappa_c})$          & (\ref{eqn:PeTildeDef})       & \renewcommand{\arraystretch}{0.7}\begin{tabular}[c]{@{}l@{}}Electromagnetic forcing\\ Péclet number\end{tabular}    
\end{tabular}
\caption{Scales and nondimensional parameters. All quantities below the dashed line are nondimensional.}
\label{tab:NdimParamDefs}
\end{center}
\end{table}

We assume vertical hydrostasy, $0={\partial_{\zeta}\Pi}$, and cyclostrophic balance, ${\upsilon_{\theta}^2}/{\rho}=\partial_{\rho}\Pi$, achieved via deflection of the free surface. Under these conditions, the meridional flow $\boldsymbol{\upsilon}_\bot = \upsilon_\rho \boldsymbol{e_r} + \mathcal{H}\upsilon_\zeta \boldsymbol{e_z}$ vanishes, leaving a linear equation for azimuthal magneto-Stokes flow:

\begin{equation}\label{eqn:PDETheta}
    \Rey \partial_{\tau}\upsilon_{\theta }={\nabla}^{2}_{\bot} \upsilon_\theta - \mathcal{H}^2\frac{\upsilon_\theta}{\rho^2}
    +\frac{\mathcal{R} +1}{\rho}\Upsilon(\tau),
\end{equation}
where we have defined the operator
\begin{equation}
    {\nabla}^{2}_{\bot}(\cdot) = \left[\mathcal{H}^2 {\rho}^{-1} \partial_{\rho}{\rho}\partial_{\rho} + \partial_{\zeta}^2\right] (\cdot)
\end{equation}

We consider the rectangular domain $(\rho,\zeta) \in (\mathcal{R},1)\times(0,1)$ with boundary conditions
\begin{equation}\label{eqn:PDEThetaBCs}
    \upsilon_\theta(\mathcal{R},\zeta,\tau)=\upsilon_\theta(1,\zeta,\tau)=\upsilon_\theta(\rho,0,\tau) = \left[\partial_{\zeta} \upsilon_\theta\right]_{\zeta=1} = 0.
\end{equation}
This approximation to the free surface boundary conditions is appropriate as long as the deflection due to capillary action and centrifugation is negligible (e.g., \citealt{greenspan_time-dependent_1963}). The Froude number $\textit{Fr}$ compares the deflection of the free surface due to centrifugation ($\sim U^2 g^{-1}$) to the depth of the fluid ($h$):
\begin{equation}\label{eqn:Frdef}
    \textit{Fr} = \sqrt{{U^2 g^{-1}}/{h}}.
\end{equation}

Surface deflection due to capillary rise or dewetting is characterised by the capillary length (e.g., \citealt{martino_surface_2006}), $l = \sqrt{\gamma/(\varrho g)}$. The Bond number $\textit{Bo}$ compares this lengthscale to the fluid depth:
\begin{equation}\label{eqn:Bodef}
    \textit{Bo} = (h/l)^2 = \varrho g h^2/\gamma.
\end{equation}

In our validation experiments, $\textit{Fr} \lesssim 10^{-2}$ and $\textit{Bo} \ge 4.4$, so we adopt (\ref{eqn:PDEThetaBCs}).

\subsection{Spin up from rest}
Solutions to (\ref{eqn:PDETheta}, \ref{eqn:PDEThetaBCs}) that develop from an initially quiescent fluid ($\upsilon_\theta=0$ at $\tau = 0$) once constant electric current is applied ($\Upsilon = 1$ for $\tau>0$) may be expressed as
\begin{equation}\label{eqn:combSoln}
    \upsilon_\theta(\rho,\zeta,\tau) = \bar{\upsilon}_\theta(\rho,\zeta) + \upsilon_\theta'(\rho,\zeta,\tau).
\end{equation}
The stationary component is
\begin{equation}\label{eqn:steadySoln}
    \bar{\upsilon}_{\theta}(\rho,\zeta)=\left(\frac{\mathcal{R}+1}{2}\right)\frac{2\zeta-\zeta^2}{\rho}-\sum_{n=1}^{\infty}\frac{2(\mathcal{R}+1)}{k_n^2 \mathcal{H}} \left[A_n \rmI_1\left(\frac{k_n }{\mathcal{H}}\rho\right)+B_n \rmK_1\left(\frac{k_n }{\mathcal{H} }\rho\right)\right] \sin (k_n  \zeta),
\end{equation}
with
\begin{subequations}\label{eqn:AnBn}
    \begin{equation}\label{eqn:An}
    A_n=\frac{\mathcal{H}}{k_n \mathcal{R}}\left[\frac{\mathcal{R}  \rmK_1\left({k_n \mathcal{R}}/{\mathcal{H} }\right)-\rmK_1\left({k_n }/{\mathcal{H}}\right)}{\rmI_1\left({k_n}/{\mathcal{H} }\right) \rmK_1\left({k_n\mathcal{R} }/{\mathcal{H}}\right)-\rmK_1\left({k_n}/{\mathcal{H}\\}\right) \rmI_1\left({k_n \mathcal{R} }/{\mathcal{H}}\right)}\right],
\end{equation}

\begin{equation}\label{eqn:Bn}
    B_n=\frac{\rmI_1\left({k_n}/{\mathcal{H}}\right)-\mathcal{R}  \rmI_1\left({k_n\mathcal{R} }/{\mathcal{H}}\right)}{\mathcal{R}  \rmK_1\left({k_n \mathcal{R}}/{\mathcal{H} }\right)-\rmK_1\left({k_n }/{\mathcal{H}}\right)} A_n,
\end{equation}
\end{subequations}
where $\rmI_1$ and $\rmK_1$ denote modified Bessel functions of the first and second kind, respectively, and $k_n=\pi (n-1/2)$.

The exact form of the time-dependent component $\upsilon_\theta'(\rho,\zeta,\tau)$ is provided in Appendix \ref{appA}, but is well-approximated for shallow layers ($\Gamma \ll 1$) by
\begin{equation}\label{eqn:1dSolnGravest}
    \upsilon_\theta'(\rho,\zeta,\tau) \approx -\bar{\upsilon}_{\theta}(\rho,\zeta)\exp\left(-\frac{T \tau}{T_\text{su}} \right),
\end{equation}
where the characteristic timescale for spin up from rest is

\begin{equation}\label{eqn:tausudef}
    T_\text{su} = \frac{4 \Rey}{\pi ^2}T = \frac{4 h^2}{\pi^2 \nu }.
\end{equation}

\subsection{Shallow and deep-layer regimes}

The first term in (\ref{eqn:steadySoln}) is equal to the asymptotic solution in the shallow-layer limit $\Gamma \to 0$:
\begin{equation}\label{eqn:shallowSoln}
    \lim_{\Gamma \to 0}\bar{\upsilon}_{\theta}=\left(\frac{\mathcal{R}+1}{2}\right)\frac{2\zeta-\zeta^2}{\rho},
\end{equation}
which inherits the inverse dependence on radius of the Lorentz force (since $\lVert \boldsymbol{J}\rVert \propto 1/r$). The surface velocity profile is then identical to Taylor-Couette flow with inner and outer sidewall rotation rates given by:
\begin{equation}
    \Omega_i = \frac{B_0 I(t) h}{4 \pi \varrho \nu r_i^2}  \quad \text{and} \quad \Omega_o = \frac{B_0 I(t) h}{4 \pi \varrho \nu r_o^2},
\end{equation}
and naturally shares Taylor-Couette flow's kinematic reversibility for $\Rey \ll 1$ \citep{taylor_film_1967}.

For $\mathcal{H}>0$, the series in (\ref{eqn:steadySoln}) produces boundary layers at both sidewalls with 95\% thicknesses $\delta_{i}$, $\delta_{o}$ defined such that
\begin{equation}\label{eqn:delta95def}
    \bar{\upsilon}_\theta = 0.95\lim_{\Gamma \to 0}\bar{\upsilon}_{\theta} \quad \text{at} \quad \rho = (r_i + \delta_{i})/r_o,\; (r_o - \delta_{o})/r_o, \quad \zeta = 1.
\end{equation}

\begin{figure}
  \centerline{\includegraphics{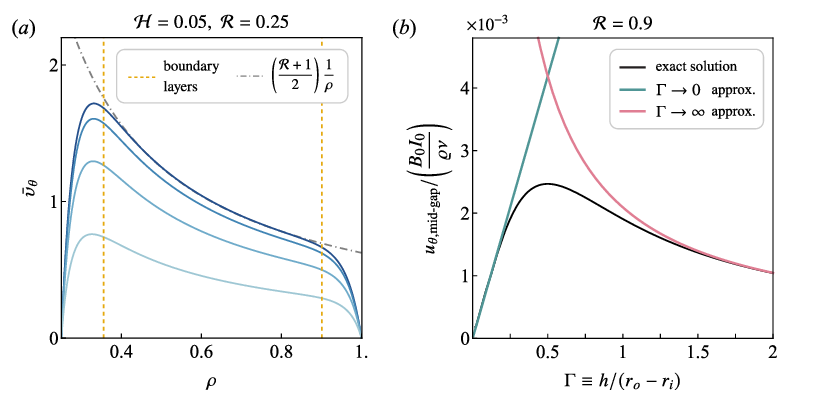}}
  \caption{(\textit{a}) Stationary magneto-Stokes flow solution given by (\ref{eqn:steadySoln}) for a channel with geometric ratios $\mathcal{H}=0.05$ and $\mathcal{R}=0.25$. Labelled contours trace the solution profile at different heights $\zeta$ above the channel base. A dash-dotted grey line shows the solution in the shallow limit ($\Gamma \to 0$), and yellow dashed lines correspond to the 95\% thicknesses of the sidewall boundary layers. (\textit{b}) Magnitude of the surface mid-gap velocity as a function of depth-to-gap-width ratio $\Gamma$ for a channel with $\mathcal{R}=0.9$. The exact solution (black) is computed using (\ref{eqn:steadySoln}). The curves corresponding to shallow (teal) and deep-layer (pink) asymptotic solutions are plotted using (\ref{eqn:shallowSoln}) and (\ref{eqn:deepsoln}), respectively.}
\label{fig:soln}
\end{figure}

Figure \ref{fig:soln}(\textit{a}) shows the stationary solution given by (\ref{eqn:steadySoln}) for a channel with geometric ratios $\mathcal{H}=0.05$ and $\mathcal{R}=0.25$. Contours correspond to different vertical positions $\zeta$ within the fluid, and dashed yellow lines indicate the 95\% thicknesses of the sidewall boundary layers. The size of each boundary layer relative to the channel width scales as
\begin{equation}\label{eqn:BLscaling}
    \Delta_i \equiv \frac{\delta_{i}}{r_o - r_i} \approx 2.51 \Gamma^{1.08}, \; \Delta_o \equiv \frac{\delta_{o}}{r_o - r_i}\approx 2.11 \Gamma^{1.06}.
\end{equation}
All numerical factors and powers in (\ref{eqn:BLscaling}) are fit to values of $\Delta_i$, $\Delta_o$ computed from (\ref{eqn:steadySoln}) for $0.01 \le \mathcal{H} \le 0.3$ and  $0.01 \le \mathcal{R} \le 0.99$. 

\begin{figure}
  \centerline{\includegraphics{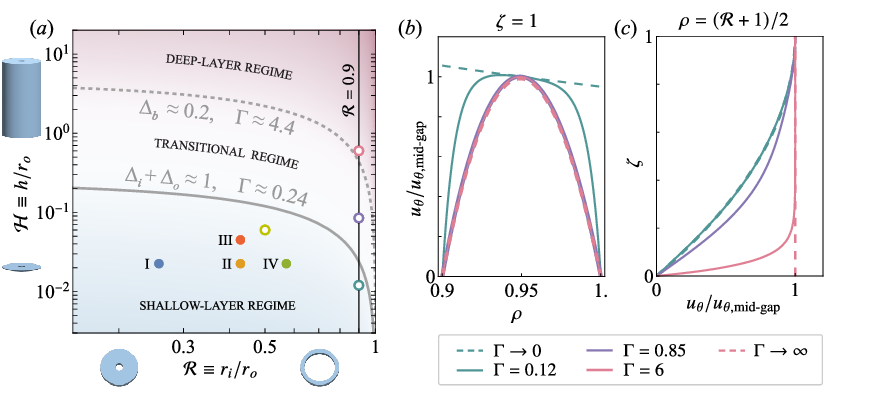}}
  \caption{(\textit{a}) Regime diagram for annular magneto-Stokes flow in the space of radius ratios $\mathcal{R}$ and aspect ratios $\mathcal{H}$. Renderings of cylindrical annuli correspond to axes values. Background tones grade from cool to warm with increasing $\Gamma$. A solid grey line indicates the boundary $\Delta_i + \Delta_o \approx 1$ (predicted by \ref{eqn:BLscaling}) between shallow-layer and transitional regimes, while a dashed grey line indicates the boundary $\Delta_b \approx 0.2$ (predicted by \ref{eqn:BLbasalscaling}) between transitional and deep-layer regimes. Points labelled with roman numerals correspond to laboratory cases discussed in \S\ref{sec:methods},\ref{sec:results}. Open markers correspond to DNS cases discussed in \S\ref{sec:apps:mixing}. (\textit{b}) A comparison of magneto-Stokes flows in channels of varying depth-to-gap-width ratio $\Gamma$ at a fixed radius ratio $\mathcal{R}=0.9$. Plotted are radial profiles of surface azimuthal velocity predicted from theory (\ref{eqn:steadySoln}) and scaled by surface values at the channel centre, $u_{\theta,\text{mid-gap}}$. Solid curves correspond to open markers of the same colour in the regime diagram (panel \textit{a}). Dashed teal and pink curves show the corresponding shallow (\ref{eqn:shallowSoln}) and deep (\ref{eqn:deepsoln}) asymptotic solutions, respectively.}
\label{fig:regimes}
\end{figure}

The scalings (\ref{eqn:BLscaling}) predict a shallow-layer regime in which sidewall boundary layers are distinct (i.e. $\Delta_i + \Delta_o < 1$) for $\Gamma \lesssim 0.24$. This regime is also characterised by the growth of the mid-gap velocity magnitude 
\begin{equation}\label{eqn:umidgapdef}
    u_{\theta,\text{mid-gap}}=U\bar{\upsilon}_\theta(\rho=\mathcal{R}/2+1/2,\zeta=1)
\end{equation}
with layer depth $h$ necessary for the basal viscous drag ($\sim 2{\nu U}/{h^2} \propto U h^{-2}$) to balance the Lorentz force ($\sim B_0 I_0/[\pi  h \varrho(r_i+r_o)] \propto h^{-1}$). Figure \ref{fig:soln}(b) shows the growth of $u_{\theta,\text{mid-gap}}$ (black curve) with layer depth closely following the linear dependence predicted by the asymptotic shallow-layer solution (\ref{eqn:shallowSoln}) (teal curve) for $\Gamma \lesssim 0.24$.

If the layer depth is increased (while still keeping inertial forces small, $\Rey \ll 1$), the dominant balance of Lorentz and sidewall viscous drag forces ($\sim 8 \nu U_\text{deep}/[r_o-r_i]^2$) leads to an alternate velocity scale
\begin{equation}\label{eqn:Udeepdef}
    U_\text{deep}= \frac{B_0 I_0 (r_o-r_i)^2}{8 \pi  h \nu  \varrho (r_i+r_o)}=\frac{U}{4 \Gamma^2}.
\end{equation}
Using $U_\text{deep}$ to nondimensionalise velocity as $w_\theta = u_\theta/U_\text{deep}$, the steady solution for an infinitely deep channel is
\begin{equation}\label{eqn:deepsoln}
    \lim_{\Gamma \to\infty} \bar{w}_\theta=\frac{2  \left(1-\rho^2\right) \mathcal{R} ^2 \log (\mathcal{R} )-2 \left(1-\mathcal{R} ^2\right) \rho^2 \log \left(\rho\right)}{(1-\mathcal{R})^3 \rho}. 
\end{equation}
See \cite{gleeson_modelling_2004} for the dimensional form of (\ref{eqn:deepsoln}).

For deep channels of finite depth ($0.24 \lesssim \Gamma < \infty$), flow is invariant with height outside of a basal boundary layer with 95\% thickness $\delta_{b}$ defined such that
\begin{equation}\label{eqn:delta95bdef}
    \bar{w}_\theta = 0.95\lim_{\Gamma \to \infty}\bar{w}_{\theta} \quad \text{at} \quad \rho = (1+\mathcal{R})/2, \quad \zeta = \delta_{b}/h.
\end{equation}
The relative thickness scales as
\begin{equation}\label{eqn:BLbasalscaling}
    \Delta_b \equiv  \delta_{b}/h \approx 0.831 \Gamma^{-0.961}.
\end{equation}
The numerical factor and power in (\ref{eqn:BLbasalscaling}) are fit to values of $\Delta_b$ computed from (\ref{eqn:steadySoln}) for $1 \le \mathcal{H} \le 20$ and  $0.01 \le \mathcal{R} \le 0.99$.

We may then define a deep-layer regime with the condition $\Delta_b < 0.2$, satisfied for $\Gamma \gtrsim 4.4$. This regime is also characterised by the decrease of  $u_{\theta,\text{mid-gap}}$ with layer depth, since the dependence of $u_{\theta,\text{mid-gap}}$ on $h$ is mainly controlled by the Lorentz force ($\propto h^{-1}$). Figure \ref{fig:soln}(b) shows the change in $u_{\theta,\text{mid-gap}}$ (black curve) with layer depth closely following the inverse dependence predicted by the asymptotic deep-layer solution (\ref{eqn:deepsoln}) (pink curve) for $\Gamma > 1$.

Figure \ref{fig:regimes}(\textit{a}) uses the conditions $\Delta_i + \Delta_o < 1$ and $\Delta_b < 0.2$ with the scalings (\ref{eqn:BLscaling}, \ref{eqn:BLbasalscaling}) to demarcate shallow and deep-layer regimes in the space of aspect and radius ratios. Teal, purple, and pink dots in figure \ref{fig:regimes}(\textit{a}) correspond to shallow-layer, transitional, and deep-layer flows in channels of the same radius ratio $\mathcal{R}=0.9$, whose predicted radial and vertical profiles are plotted in panels (\textit{b}) and (\textit{c}), respectively. Asymptotic shallow and deep-layer solutions (\ref{eqn:shallowSoln}, \ref{eqn:deepsoln}) are plotted as dashed curves.
\bigskip
\section{Experimental methods}\label{sec:methods}
We validate the approximate solution (\ref{eqn:combSoln}) via four laboratory experiments (Cases I–IV), matched with direct numerical simulations (DNS) of the nonlinear axisymmetric flow governed by (\ref{eqn:3Dgov}, \ref{eqn:3Dcont}). This complementary approach permits us to test various physical effects not accounted for in our model: the DNS test the impact of meridional flow alone, while the laboratory experiments add the effects of surface tension and a dynamic free surface. The results of these validation cases are discussed in \S\ref{sec:results}. An additional laboratory experiment (Case HS) is detailed in \S\ref{sec:apps}.

\subsection{Laboratory experiments}\label{sec:methods:lab}
Validation experiments (Cases I–IV) are performed using an open-top annular channel consisting of a 17.5 cm-radius steel outer cylindrical sidewall, acrylic base, and a removable stainless steel inner cylindrical sidewall, which may be replaced with cylinders of different radii. The channel is placed inside the solenoidal electromagnet bore of UCLA's RoMag device \citep{xu_thermoelectric_2022}, and a DC bench power supply provides a controlled current $I$ between the channel sidewalls. An $80.0 \pm 0.05$ g/L NaCl:H$_2$O solution is used as the working fluid for all cases; 0.1 mL of dish detergent is added for every litre of solution to reduce surface tension, which is measured as $\gamma = 38 \pm 4$ mN/m using the capillary rise method (e.g., \citealt{martino_surface_2006}). The fluid is estimated to have electrical conductivity $\sigma = 12.3 \pm 0.1$ S/m, kinematic viscosity $\nu = (1.10 \pm 0.05) \times 10^{-6}$ m$^2$/s, and density $\varrho = 1059.1 \pm 0.7$ kg/m$^3$, using the salinity-based models of \cite{park_partial_1964}, \cite{isdale_physical_1972}, and \cite{isdale_physical_1972-1}, respectively.

\begin{table}
\begin{center}
\def~{\hphantom{0}}
\addtolength{\tabcolsep}{1pt} 
\begin{tabular}{ccccccc}
\begin{tabular}[c]{@{}c@{}}Case\\{} \end{tabular} & \begin{tabular}[c]{@{}c@{}}$I_0$ (\textit{a})\\ ($\pm 0.003$ A)\end{tabular} & \begin{tabular}[c]{@{}c@{}}$B_0$ (mT)\\{}\end{tabular} & \begin{tabular}[c]{@{}c@{}}$h$ (cm)\\ ($\pm 0.05$ cm)\end{tabular} & \begin{tabular}[c]{@{}c@{}}$r_i$ (cm)\\ ($\pm 0.03$ cm)\end{tabular} & \begin{tabular}[c]{@{}c@{}}$r_o$ (cm)\\ ($\pm 0.03$ cm)\end{tabular} & \begin{tabular}[c]{@{}c@{}}$U$ (cm/s)\\{}\end{tabular} \\ \hline
I      & 0.081     & $30.0 \pm 0.2$  & 0.4      & ~4.44      & 17.52      & $0.56 \pm 0.08$ \\
II     & 0.099     & $30.0 \pm 0.6$  & 0.4      & ~7.62      & 17.52      & $0.63 \pm 0.09$ \\
III    & 0.039     & $20.0 \pm 0.4$  & 0.8      & ~7.62      & 17.52      & $0.36 \pm 0.04$ \\
IV     & 0.099     & $35~ \pm 1~$  & 0.4      & 10.14      & 17.52      & $0.7 \pm 0.1$   \\
HS     & 0.060     & $190 \pm 1~$  & 0.2      & ~3.75      & ~9.84      & $0.24 \pm 0.06$ \\
\end{tabular}
\caption{Dimensional experimental parameters and predicted velocity $U$ at surface mid-gap ($z=h$, $r=[r_i+r_o]/2$), computed using (\ref{eq:UMS}). Error in values of $U$ reflect the propagation of measurement uncertainty of the control parameters.}
\label{tab:Dim}
\end{center}
\end{table}

Inner radius, fluid height, electric current, and magnetic field strength are varied across Cases I–IV; values of these dimensional parameters are reported in table \ref{tab:Dim}. Cases I–IV span three different radius ratios $\mathcal{R}$ = 0.25, 0.44, 0.58 and two different aspect ratios $\mathcal{H}$ = 0.023, 0.046; these values correspond to the solid points in Figure \ref{fig:regimes}(\textit{a}). Fluid depth is kept above the capillary length $l = 0.19 \pm 0.01 $ cm ($\textit{Bo} > 1$) to minimise relative differences in depth due to capillary action. Voltage across the electrodes is maintained under $\sim$1.5 V to prevent electrolysis of water. Under this constraint, electric current and magnetic field strength are held between 0.04–0.1 A and 20–35 mT, respectively, to keep $\Rey < 1$ and $\textit{Fr}^2 \ll 1$. Values of these nondimensional control parameters are reported in table \ref{tab:Ndim}.

\begin{table}
\begin{center}
\def~{\hphantom{0}}
\addtolength{\tabcolsep}{4pt}
\renewcommand{\arraystretch}{0.5}
\begin{tabular}{cccccccccc}
Case   & $\mathcal{R}$ & $\mathcal{H}$ & $\Rey$  & $\textit{Fr}$ & $\textit{Bo}$ & $\textit{Ha}$       & $\textit{Rm}$        \\ \hline
I      & 0.25          & 0.023         & 0.38~   & 0.030         & 4.4           & 0.012               & $6.6\times 10^{-12}$ \\
II     & 0.44          & 0.023         & 0.36~   & 0.032         & 4.4           & 0.012               & $6.3\times 10^{-12}$ \\
III    & 0.44          & 0.046         & 0.78~   & 0.012         & 18            & 0.016               & $1.3\times 10^{-11}$ \\
IV     & 0.58          & 0.023         & 0.35~   & 0.034         & 4.4           & 0.014               & $6.1\times 10^{-12}$ \\
HS     & 0.38          & 0.020         & 0.058   & 0.016         & 1.1           & $3.8\times 10^{-3}$ & $1.0\times 10^{-12}$ \\
\end{tabular}
\caption{Nondimensional experimental parameters. The radius ratio $\mathcal{R}$, aspect ratio $\mathcal{H}$, control Reynolds number $\Rey$, Froude number $\textit{Fr}$, Bond number $\textit{Bo}$, Hartmann number $\textit{Ha}$, and magnetic Reynolds number  $\textit{Rm}$ are defined in \S\ref{sec:theory}.}
\label{tab:Ndim}
\end{center}
\end{table}

Before the start of each case (I–IV), a streak of buoyant blue dye is drawn across the quiescent fluid surface. From $t=0$ to $t=0.5 \pi T$, the power supply drives a constant current $I_0$, and an overhead camera records the dye advection. Blue-channel thresholding and Canny edge detection \citep{canny_computational_1986,bradski_opencv_2000} are applied to the perspective-corrected video in order to track the (Lagrangian) angular position $\theta(r,t)$ of the leading dye streak edge. Uncertainty in $\theta(r,t)$ is computed as the change in estimated position under a 10\% adjustment of color threshold values. At every $\Delta t = 0.5 T_\text{su}$, $\theta$ is determined at 15 points across the channel (in $r$), excluding the menisci at the sidewalls where dye spreads rapidly via adhesion instead of advection. A time-series of surface velocity $u_\theta(r)$ is then estimated via second-order central difference of $\theta(r,t)$ over time.

\subsection{Direct numerical simulations}\label{sec:methods:DNS}
Cases I–IV are matched with DNS of nonlinear, axisymmetric flow governed by (\ref{eqn:3Dgov}, \ref{eqn:3Dcont}) with initially quiescent flow ($\boldsymbol{\upsilon}=0$ at $\tau = 0$, $\Upsilon = 1$ for $\tau>0$) and no-slip conditions on all boundaries except for the surface, which is treated as a free-slip rigid lid: $\boldsymbol{\upsilon}(\mathcal{R},\zeta,\tau)=\boldsymbol{\upsilon}(1,\zeta,\tau)=\boldsymbol{\upsilon}(\rho,0,\tau) = 0 $, $\left[\partial_{\zeta} \upsilon_\rho\right]_{\zeta=1}=\left[\partial_{\zeta} \upsilon_\theta\right]_{\zeta=1}=\upsilon_\zeta(\rho,1,\tau) = 0$. We employ the Dedalus pseudospectral framework \citep{burns_dedalus_2020}, 
expanding $\upsilon_\zeta$ in 512 sine modes in $\zeta$ and all other fields in 512 cosine modes in $\zeta$; all fields are expanded in $\rho$ with 256 Chebyshev modes. The no-slip condition is enforced at $\zeta=0$  using the volume-penalty method \citep{hester_improving_2021}, which adds spatially-masked linear damping terms $-G({\zeta}/{\delta}){\upsilon_\rho}/{\tau_{\text{VP}}}$, $-G({\zeta}/{\delta}){\upsilon_\theta}/{\tau_{\text{VP}}}$, $-G({\zeta}/{\delta}){\upsilon_\zeta}/{\tau_{\text{VP}}}$ to the corresponding components of the axisymmetric momentum equation (\ref{eqn:3Dgov}),
where $G(x) = [1-\rm{erf}(\sqrt{\pi}x)]/2$ is a masking function and $\tau_{\text{VP}}$ is the volume-penalty damping timescale nondimensionalised by $T$. The masking function is smoothed over a vertical length scale $\delta$ (set here to $\delta=0.01$), which is used to determine an appropriate value for $\tau_{\text{VP}}$. This is effected by requiring the damping length-scale $\sqrt{\tau_{\text{VP}}/\Rey}$ to be proportional to the smoothing scale: $\sqrt{\tau_{\text{VP}}/\Rey} =\delta/\delta^*$. The proportionality constant $\delta^*$ is set to the optimal value found by \cite{hester_improving_2021}, $\delta^*=3.11346786$; this choice of $\delta^*$ eliminates the displacement length error associated with the mask, $G(x)$. See \cite{hester_improving_2021} for details on optimising the volume-penalty method.

In all simulations, the system is integrated from $\tau=0$ to $\tau = 1.1\pi$ over $10^4$ time steps using the second-order semi-implicit backwards difference (SBDF2) scheme (\citealt{wang_variable_2008}, Eqn. 2.8). Acceleration, pressure, and rectilinear viscous terms are time-integrated implicitly; the remaining terms are treated explicitly.

Reported results match those obtained at half their spatial resolution as well as the analytical solution at $t= 3 T_\text{su}$. The code used for these simulations and for the dye-tracking described in \S\ref{sec:methods:lab} is available online (\url{https://github.com/cysdavid/magnetoStokes}). 
\section{Results}\label{sec:results}

\begin{figure}
  \centerline{\includegraphics{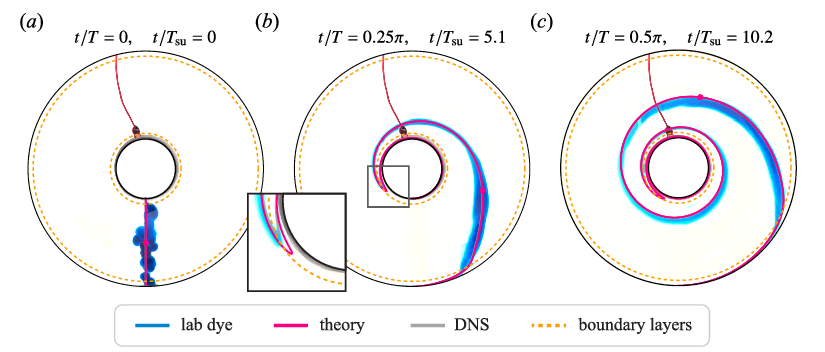}}
  \caption{Snapshots of a free surface dye track (blue) from laboratory Case I, (\textit{a}) when power is turned on, (\textit{b}) after $\sim$5 spin up times, and (\textit{c}) after $\sim$10 spin up times. Time integrations of the approximate analytical solution (\ref{eqn:combSoln}) and DNS result are overlain as magenta and grey curves, respectively. Dotted yellow circles correspond to the 95\% thickness of each sidewall boundary layer as predicted from (\ref{eqn:combSoln}). The red cord near the 12 o'clock position in each photograph is the electrical wire leading from the power supply to the inner electrode.}
\label{fig:spiral}
\end{figure}

Figure \ref{fig:spiral} shows three photographs of laboratory Case I, taken when power is turned on (\textit{a}), after 0.25 circulation times (\textit{b}), and after 0.5 circulation times (\textit{c}). The time-integrated analytical solution (\ref{eqn:combSoln}) and DNS results are overlain in magenta and grey, respectively. Laboratory, analytical, and numerical results match well in the bulk, differing most within the boundary layers (dashed yellow lines). A closeup (\textit{b}, inset) of the inner boundary layer shows the laboratory dye streak and DNS curve trailing behind the analytical solution, resulting from the  parasitic effect of meridional circulation on the steady-state azimuthal flow and from a lag in spin-up. The bulk flow also exhibits finite spin-up time effects. In each panel of figure \ref{fig:spiral}, a magenta dot is placed on the analytical curve at $r = (r_i + r_o)/2$. For flow that has fully spun up, $\pi T$ is the time it takes for this dot to make one revolution. In figure \ref{fig:spiral}(\textit{b}), the magenta dot has travelled slightly less than a quarter revolution from $t=0$ to $t = 0.25 \pi T$, a product of the finite $\Rey$ value in our experiments.

\begin{figure}
  \centerline{\includegraphics{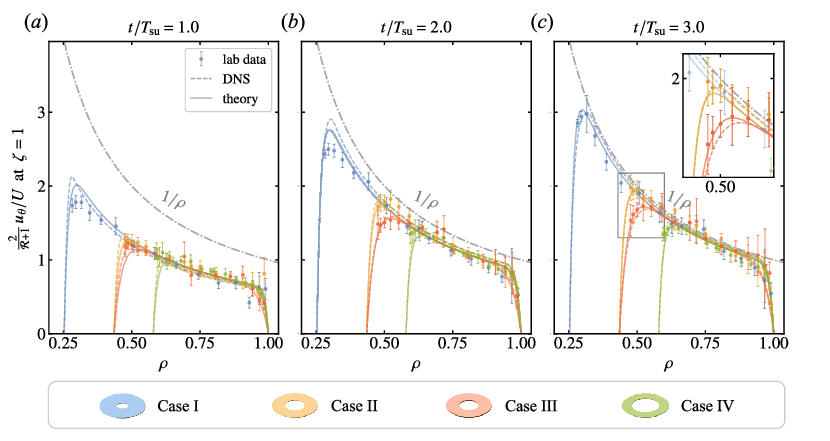}}
  \caption{Radial profiles of scaled azimuthal velocity at the free surface for Cases I-IV at (\textit{a}) 1 spin-up time after rest, (\textit{b}) 2 spin-up times, and (\textit{c}) 3 spin-up times. Solid theoretical curves are computed using (\ref{eqn:combSoln}). DNS are shown via dashed curves. Also plotted is the $1/\rho$ profile (dash-dotted grey curve) corresponding to the shallow ($\Gamma \to 0$), long-time solution (\ref{eqn:shallowSoln}) at $\zeta=1$. Error bars represent $\pm 1$ standard deviation propagated from uncertainty in the dye tracking velocimetry algorithm and from uncertainty in the predicted velocity scale $U$ used to normalise the data}. Rendered cylindrical annuli in the lower legend depict the channel geometry of each case.
\label{fig:profs}
\end{figure}

Figure \ref{fig:profs} plots radial profiles of scaled azimuthal velocity for all four cases. Included are laboratory data (points), DNS results (dashed curves), and the approximate analytical solution given by (\ref{eqn:combSoln}) (solid curves) after 1 spin-up time (\textit{a}), 2 spin-up times (\textit{b}), and 3 spin-up times (\textit{c}). Bars on the data points correspond to error introduced by the dye-tracking algorithm and from measurement uncertainty propagated through the computed velocity scale $U$. The scaled velocity profiles evolve towards the $1/\rho$ curve (gray dash-dotted line) over time, apart from the boundary layers. A closeup of the inner boundary layer for the two cases with $\mathcal{R} = 0.44$ (panel \textit{c}, inset) shows a clear separation between the profiles for $\mathcal{H} = 0.023$ (Case II) and $\mathcal{H} = 0.046$ (Case III) and excellent agreement with theory.

A slight separation between data, DNS, and theory at Case I's peak in velocity (near $\rho=0.3$) for $t/T_\text{su} \le 2$ can be seen in figure \ref{fig:profs}(\textit{a,b}). The laboratory flow (blue points) spins up slower than the approximate solution (solid blue line), while the DNS result (dashed blue line) spins up faster. The gap between laboratory data and theory may be related to the dynamic adjustment of the free surface or the reduction of current density at sidewall menisci. In contrast, the gap between theory and DNS is largely the result of the sidewall viscous drag's effect on spin-up. As evident in the full solution in Appendix \ref{appA}, features with higher spatial frequency in the radial direction (e.g., the sharp peak near $\rho=0.3$) spin up faster than lower frequency features. This effect is neglected in the approximate solution (\ref{eqn:combSoln}) but is retained in the DNS.

These finite-$\Rey$ simulations also retain nonlinear advection, and each DNS case exhibits a steady (for $t \gg T_{\text{su}}$) clockwise vortex in the $\rho,\zeta$-plane near the inner sidewall where centrifugal forces are strongest (cf. \citealt{norouzi_analytical_2013}). This meridional circulation has a parasitic effect on the azimuthal flow, resulting in the slight gap between DNS and theory in figure \ref{fig:profs}(\textit{c}). 
The largest root mean square error between theory and DNS (Case I) is  3.3\% of the average DNS azimuthal velocity at $t=3 T_{\text{su}}$. This difference is smaller than the error bars on the laboratory data and is expected to vanish with decreasing $\Rey$.

\section{Low-\texorpdfstring{$\Rey$}{Re} mixing in magneto-Stokes flow}\label{sec:apps}

The expansion of ``lab on a chip" technology across a range of industrial and biomedical fields has increased demand for microfluidic devices that can efficiently mix chemical species at low $\Rey$ \citep{pamme_magnetism_2006,mansur_state---art_2008,jeong_applications_2010}. Magneto-Stokes systems \citep{yi_magnetohydrodynamic_2002,gleeson_modelling_2004,west_structuring_2003} are particularly well-suited to this purpose because they have no moving components that require miniaturisation and they work in simple, easily-fabricated channel geometries (cf. \citealt{ehrfeld_characterization_1999,bertsch_static_2001}).

In the following subsections, we analyse the properties of magneto-Stokes flow relevant to low-$\Rey$ mixing. The design of efficient micromixers often focuses on the generation of vortices (\citealt{sudarsan_multivortex_2006, chang_electrokinetic_2007} and ref. therein), which tend to augment mixing (e.g., \citealt{cetegen_experiments_1993}). Therefore, in subsection \ref{sec:apps:HeleShaw}, we determine the conditions under which the Lorentz force can produce vorticity in shallow-layer magneto-Stokes flows in arbitrary (2D) channel geometry. In micromixers that drive flow via electroosmosis rather than MHD, the generation of vorticity hinges on breaking the similitude between velocity and the electric field \citep{cummings_conditions_2000}. An analogous similitude property exists for many magneto-Stokes flows, including our annular configuration, for which the 2D velocity field and the Lorentz force are everywhere proportional by the same amount. In contrast to electroosmotic flows, annular magneto-Stokes flow is irrotational even when obstacles are placed in the channel to break similitude (cf. the obstacle-induced electroosmotic vortices in \citealt{dukhin_electrokinetic_1991,ben_nonlinear_2002}).

We show in subsection \ref{sec:apps:mixing} that, despite the lack of significant axial vorticity, shallow-layer annular magneto-Stokes flow enhances mixing via Taylor-dispersion \citep{taylor_dispersion_1953,aris_dispersion_1956} or through an advectively-dominated mechanism similar to that of a point-vortex flow \citep{rhines_how_1983,flohr_accelerated_1997}. Our results extend to transitional and deep-layer flows, and they demonstrate that shear-enhancement of mixing is initiated for the least electromagnetic effort ($B_0 I_0$) in $\Gamma \approx 1$ channels.

\subsection{Irrotationality in shallow-layer magneto-Stokes systems}\label{sec:apps:HeleShaw}

We consider a magneto-Stokes micromixer of uniform depth $h$ and arbitrary planform (i.e., lateral boundary geometry) placed in a vertical magnetic field $\boldsymbol{B} = B_z \boldsymbol{e_z}$. The governing equation (\ref{eqn:PDETheta}) generalises to
\begin{equation}\label{eqn:genchannelPDE}
    \Rey \left(\partial_\tau \boldsymbol{\upsilon} + \boldsymbol{\upsilon}\bcdot\bnabla_{\text{2D}}\boldsymbol{\upsilon}\right)=-\bnabla_{\text{2D}}P + \mathcal{H}^2\nabla_{\text{2D}}^2\boldsymbol{\upsilon} + \partial_\zeta^2\boldsymbol{\upsilon} + \boldsymbol{f}, \quad \bnabla_{\text{2D}} \bcdot \boldsymbol{\upsilon} = 0,
\end{equation}
where we now permit nonaxisymmetry and lateral pressure gradients but retain vertical hydrostasy ($\boldsymbol{\upsilon}
\bcdot \boldsymbol{e_z} = 0$). Here, we have altered the nondimensionalisation in \S\ref{sec:theory:gov} to use a generic horizontal lengthscale $L$ in place of $r_o$, $r_i$, and defined $\bnabla_{\text{2D}}$ such that $\bnabla (\cdot)= h^{-1}\left[\mathcal{H}\bnabla_{\text{2D}} + \boldsymbol{e_z}\partial_\zeta\right](\cdot)$. Dimensionless pressure $P$ and horizontal Lorentz force $\boldsymbol{f}$ have been scaled with $\varrho \nu U L h^{-2}$ and $\varrho \nu U h^{-2}$, respectively.

In the limits $\Rey,\mathcal{H} \to 0$, appropriate for lab-on-a-chip systems, we make a Hele-Shaw approximation \citep{hele-shaw_flow_1898} motivated by the form of the annular shallow-layer solution (\ref{eqn:shallowSoln}): $\boldsymbol{\upsilon} = (2\zeta - \zeta^2)\boldsymbol{\upsilon}_{\text{2D}}$ where $\partial_\zeta \boldsymbol{\upsilon}_{\text{2D}} = 0$. Applying these assumptions to  (\ref{eqn:genchannelPDE}) yields
\begin{equation}\label{eqn:HeleShaw}
    \boldsymbol{\upsilon}_{\text{2D}} = \frac{1}{2}\left(\boldsymbol{f}-\bnabla_{\text{2D}}P\right), \quad \boldsymbol{n} \bcdot \boldsymbol{\upsilon}_{\text{2D}} = 0 \; \text{on} \; \partial\mathcal{D},
\end{equation}
where $\boldsymbol{n}$ denotes the unit vector normal to the lateral boundaries $\partial \mathcal{D}$. Then, the quasi-2D flow is irrotational if and only if
\begin{equation}\label{eqn:irrot}
    \boldsymbol{e_z}\bcdot\left(\bnabla_{\text{2D}}\times \boldsymbol{f}\right) = \frac{h^2 \sigma}{\varrho \nu U}\left[ B_z \partial_z E_z -  \left(\boldsymbol{E}\bcdot\bnabla \right)B_z\right] = 0,
\end{equation}
where we have used Gauss' law ($\partial_z B_z = 0$) and neglected free charges ($\bnabla \bcdot \boldsymbol{E}=0$). Thus, vorticity can be generated in a magneto-Stokes micromixer given strong vertical gradients in $E_z$ or horizontal gradients in $B_z$.

Since the quasistatic electromagnetic fields can be prescribed, the resulting 2D flow may be readily predicted using (\ref{eqn:HeleShaw}) after solving the pressure equation,
\begin{equation}\label{eqn:HSpoisson}
    \nabla_{\text{2D}}^2 P = \bnabla_{\text{2D}} \bcdot \boldsymbol{f}, \quad \boldsymbol{n} \bcdot \bnabla_{\text{2D}}P = \boldsymbol{n}\bcdot \boldsymbol{f} \; \text{on} \; \partial\mathcal{D}.
\end{equation}
This problem is greatly simplified if the Lorentz force is nondivergent:
\begin{equation}\label{eqn:nondiv}
    \bnabla_{\text{2D}} \bcdot \boldsymbol{f} = \frac{h^2 \sigma}{\varrho \nu U} \boldsymbol{E}\bcdot\left(\bnabla \times \boldsymbol{B}\right)=0,
\end{equation}
and if the boundaries are perfectly-conducting:
\begin{equation}\label{eqn:perfcond}
    \boldsymbol{n}\times \boldsymbol{E}=0\; \text{on} \;\partial \mathcal{D},
\end{equation}
such that $\boldsymbol{n}\bcdot \boldsymbol{f}=0$ on $\partial \mathcal{D}$. The relations (\ref{eqn:HSpoisson}–\ref{eqn:perfcond}) imply $\bnabla_{\text{2D}} P = 0$ identically, which results in similitude between the prescribed Lorentz force and the resultant velocity field: $\boldsymbol{\upsilon}_{\text{2D}} =\frac{1}{2} \boldsymbol{f}$.

An analogous similitude exists for some electrokinetic flows, where the velocity is proportional to the applied electric field instead of the Lorentz force \citep{cummings_conditions_2000}. These electrokinetic flows are necessarily irrotational by nature of the quasistatic electric fields used to drive them, and thus the key to generating vorticity lies in breaking similitude \citep{dukhin_electrokinetic_1991,ben_nonlinear_2002}. In contrast, 2D magneto-Stokes flows are only irrotational if (\ref{eqn:irrot}) is satisfied, which is independent of the similitude conditions (\ref{eqn:nondiv}, \ref{eqn:perfcond}). Thus, magneto-Stokes micromixers can benefit simultaneously from the presence of vorticity and the analytical convenience of similitude between the velocity and imposed Lorentz force.

\begin{figure}
    \centering
    \includegraphics[width=1\linewidth]{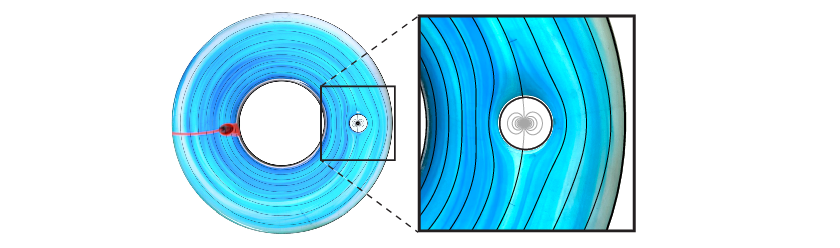}
    \caption{Dye-visualised laboratory flow (Case HS) around a circular, electrically-insulating obstacle (white disk). Overlain in black are approximate potential flow streamlines obtained using the Milne-Thomson circle theorem \citep{milne-thomson_theoretical_1938}. Gray curves indicate the potential flow doublet that produces the circular obstacle streamline.}
    \label{fig:hs}
\end{figure}

If (\ref{eqn:irrot}), (\ref{eqn:nondiv}), and (\ref{eqn:perfcond}) are all satisfied, as is the case for the annular device considered in this work, then the Lorentz force and velocity are proportional to the gradient of a potential $\Phi$ that is possibly multi-valued (like that of a point vortex; \citealt{lamb_hydrodynamics_1906}). Potential flow is maintained even when similitude is broken (e.g., by the addition of an electrically-insulating obstacle to the flow). The only change is the contribution of pressure to the total potential:
\begin{equation}
    \boldsymbol{\upsilon}_{\text{2D}} = \frac{1}{2}\bnabla_{\text{2D}}\left(\Phi-P\right).
\end{equation}
The pressure field $P$ may be constructed to enforce the no-flux condition on the lateral boundaries using potential theory (e.g., \citealt{lamb_hydrodynamics_1906,milne-thomson_theoretical_1938}).

In practice, these magneto-Stokes potential flows readily arise even with moderate gradients in magnetic field strength and fringing of electric field lines near menisci. Figure \ref{fig:hs} shows magneto-Stokes flow around an electrically-insulating plastic obstacle in the annular device (Case HS), resting on an array of permanent magnets with moderate horizontal variability in field strength (with 1 standard deviation equal to 27\% of the mean). Despite these gradients in $B_z$ and the presence of non-negligible surface tension effects ($\textit{Bo} = 1.1$), the dye streaks coincide with approximate potential flow streamlines (black overlay) obtained using the Milne-Thomson circle theorem \citep{milne-thomson_theoretical_1938}. Thus, we expect magneto-Stokes micromixers to be robust to surface tension effects and moderate variations in magnetic field strength.

\subsection{Enhanced mixing in annular magneto-Stokes flow}\label{sec:apps:mixing}
Though annular magneto-Stokes flows are vorticity-free in the shallow limit, they make for robust, easily-fabricated micromixing systems. Further, they exhibit multiple regimes of enhanced mixing, which we characterise here. Mixing effects in annular magneto-Stokes flows were first studied by \cite{gleeson_magnetohydrodynamic_2002, gleeson_modelling_2004}. The authors focused on the deep limit ($\Gamma \to \infty$), which renders the flow 2D and enabled them to derive analytical asymptotic predictions for mixing time. These scaling laws are extensively supported by 2D DNS \citep{gleeson_modelling_2004}, but exhibit large errors when compared to experimental results for the shallow-layer systems most relevant to compact, lab-on-a-chip applications \citep{west_structuring_2003}. To address this gap, we generalise the scaling laws of \cite{gleeson_modelling_2004} to finite $\Gamma$ systems, using the analytical solution developed in \S\ref{sec:theory} and validated in \S\ref{sec:results}.

The homogenisation of solute concentration $c$ is governed by
\begin{equation}\label{eqn:advdiffndm}
    \textit{Pe}\left[\partial_\tau c + (1+\mathcal{R})\omega\partial_\theta c\right] = \frac{\mathcal{H}^2}{\rho^2}\partial_\theta^2 c + {\nabla}^{2}_{\bot}c,
\end{equation}
where $\omega$ is the angular velocity field. The dominance of advection over diffusion is controlled by the Péclet number ($\textit{Pe}$), defined via the Reynolds and Schmidt ($\textit{Sc}$) numbers as
\begin{equation}\label{eqn:PeDef}
    \textit{Pe} = \textit{Sc} \Rey, \quad \textit{Sc}=\nu/\kappa_c,
\end{equation}
where $\kappa_c$ denotes the molecular diffusivity of the solute. So long as $\Rey \ll \min(\textit{Pe},1)$, the spin-up period $T_{\text{su}}$ is the shortest timescale, and we consider the flow to be quasi-steady such that the angular velocity in (\ref{eqn:advdiffndm}) may be computed as  $\omega = \bar{\upsilon}_\theta(\rho,\zeta)/\rho$ using the stationary solution (\ref{eqn:steadySoln}).

We consider a simple non-axisymmetric initial condition
\begin{equation}\label{eqn:IC}
c(\rho,\theta,\zeta,0)=c_0(\theta) =
    \begin{cases}
        0, & -\pi/2 < \theta \le \pi/2\\
        1, & \pi/2 < \theta \le 3 \pi/2
    \end{cases},
\end{equation}
and define a mixing norm $m$, following \cite{gleeson_modelling_2004}, as the normalised root mean square deviation of the concentration field from its average value, $\overline{c}$:
\begin{equation}
    m(t/T) = \frac{\left\Vert c - \overline{c}\right\Vert}{\left\Vert
c_0 - \overline{c} \right\Vert}, \quad \Vert
\cdot \Vert^2 = \frac{1}{\pi(1-\mathcal{R}^2)}\int_0^1\int_0^{2\pi}\int_{\mathcal{R}}^{1}(\cdot)^2\rho\mathrm{d}\rho\mathrm{d}\theta\mathrm{d}\zeta,
\end{equation}
such that $m(0)=1$. The mixing time $t_M$ is then defined as the time it takes for $m$ to shrink to some value $M<1$. Though other metrics exist (e.g., the eigenvalue-base approach of \citealt{cerbelli_spectral_2009}), $t_M$ benefits from its clear physical meaning and applicability to all mixing regimes. In each of these regimes, predictions for $t_M/T$ as a function of $\textit{Pe}$, $\Gamma$, $\mathcal{R}$ follow from the appropriate asymptotic reduction of (\ref{eqn:advdiffndm}).

\subsubsection{Diffusion-dominated regime}
For $\textit{Pe} \ll 1$, lateral diffusion occurs before advection can effectively shear the tracer concentration front. Solving (\ref{eqn:advdiffndm}, \ref{eqn:IC}) in the absence of advection and retaining the effect of the fundamental mode yields the scaling law
\begin{equation}\label{eqn:approxdiffscaling}
    t_M/T \sim \frac{1}{\mathcal{H}^2 \lambda_{11}^2}\ln{\left(\frac{2\sqrt{2}}{\pi M}\right)}\textit{Pe},
\end{equation}
where $\lambda_{11}$ is the smallest positive root of $\rmJ_1'(\lambda \mathcal{R})\rmY_1'(\lambda)-\rmY_1'(\lambda \mathcal{R})\rmJ_1'(\lambda)=0$. We assume $\mathcal{R} \gtrsim 0.1$ in (\ref{eqn:approxdiffscaling}) so that we may approximate an additional numerical factor arising from the average of the first radial eigenfunction with unity. Dimensionally, $t_M \sim \lambda_{11}^{-2} \ln[2\sqrt{2}/(\pi M)] r_o^2/\kappa_c$ and the mixing time is independent of depth $h$, as the problem is essentially 2D.

\subsubsection{Taylor dispersion regime}
Depth is important at intermediate values of $\textit{Pe}$, where vertical and radial shear enables rapid transverse diffusion in narrow channels. A centre-manifold reduction \citep{mercer_complete_1994,roberts_low-dimensional_1996,ding_determinism_2022,ding_diffusion-driven_2023} of (\ref{eqn:advdiffndm}) yields the scaling law
\begin{equation}\label{eqn:TDscaling}
    t_M/T \sim \frac{(1+\mathcal{R})^2}{4\;\mathcal{C}_D}\ln\left(\frac{2\sqrt{2}}{\pi M}\right)\textit{Pe}^{-1}.
\end{equation}
See Appendix \ref{app:B} for the details of this derivation. This inverse Péclet number dependence, typical of Taylor dispersion \citep{taylor_dispersion_1953}, results from an effective diffusivity $\kappa_e= \mathcal{C}_D (\textit{Pe}/\mathcal{H})^{2}{\kappa_c}$ that actually increases with the vigour of advection ($\textit{Pe}$). The dispersion coefficient $\mathcal{C}_D$ is given by
\begin{equation}\label{eqn:TaylorDispCoeff}
    \mathcal{C}_D = -\frac{(1+\mathcal{R})^2}{2}\langle{(\omega-\langle {\omega} \rangle)a_1}\rangle, \quad \langle \cdot \rangle =\frac{2}{1-\mathcal{R}^2}\int_0^1 \int_{\mathcal{R}}^1 (\cdot) \rho \mathrm{d}\rho\mathrm{d}\zeta,
\end{equation}
where the function $a_1(\rho,\zeta)$ is found by solving
\begin{equation}\label{eqn:a1_2D}
    {\nabla}^{2}_{\bot} a_1 = \frac{(1+\mathcal{R})^2}{2}(\omega-\langle {\omega} \rangle),
\end{equation}
subject to no-flux boundary conditions.

The transition between diffusion-dominated and Taylor dispersion regimes marks the onset of mixing enhancement, which occurs near the Péclet number $\textit{Pe}_0$ at which scalings (\ref{eqn:approxdiffscaling}) and (\ref{eqn:TDscaling}) are equal:
\begin{equation}\label{eqn:PeTrans}
    \textit{Pe}_0 \sim \frac{\lambda_{11}\mathcal{H}(1+\mathcal{R})}{2\;\sqrt{\mathcal{C}_D}}.
\end{equation}

\subsubsection{Advection-dominated regime}
At much higher values of $\textit{Pe}$, advection occurs rapidly enough to shear the tracer concentration front into radially and vertically interleaved lamellae. Accordingly, we transform (\ref{eqn:advdiffndm}) into the Lagrangian frame following the flow. For $\textit{Pe} \gg 1$, we recover the advection-dominated scaling
\begin{equation}\label{eqn:advscaling}
    \tau_M \sim  F^{-1}\left(\frac{\pi M}{2\sqrt{2}}\right)\textit{Pe}^{1/3},
\end{equation}
which contains the one third dependence on Péclet number found in vortex flows \citep{rhines_how_1983,flohr_accelerated_1997}. The function
\begin{equation}\label{eqn:Fadvfxn}
    F(x)=\left\Vert e^{-\frac{1}{3}(1+\mathcal{R})^2\left( {\bnabla}_\bot\omega \cdot {\bnabla}_\bot\omega\right) x^3}\right\Vert, \quad \text{where} \quad {\bnabla}_\bot(\cdot) = \left[\boldsymbol{e_r}\mathcal{H}\partial_{\rho} + \boldsymbol{e_z}\partial_{\zeta}\right](\cdot),
\end{equation}
captures the effect of shear (${\bnabla}_\bot\omega$) on advection-dominated mixing. (For details of the derivation, see Appendix \ref{app:B}.)

A Mathematica notebook, available at \url{https://github.com/cysdavid/magnetoStokes}, implements all three scaling laws (\ref{eqn:approxdiffscaling}, \ref{eqn:TDscaling}, \ref{eqn:advscaling}) as a tool for practitioners, inverting (\ref{eqn:Fadvfxn}) numerically and solving (\ref{eqn:a1_2D}) with finite elements.

\subsubsection{Comparison to 3D DNS}
\begin{figure}
    \centering
    \includegraphics[width=1\linewidth]{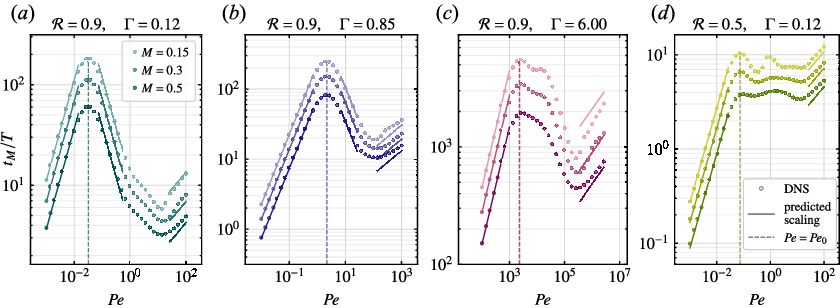}
    \caption{Mixing time $t_M$ versus Péclet number $\textit{Pe}$ for four DNS surveys in different channel geometries ($\mathcal{R}$, $\Gamma$). Results for mixing levels $M=0.5, 0.3, 0.15$ are indicated with darker to lighter tones. For each $M$, solid lines in the corresponding shade indicate predicted scaling exponents and coefficients using (\ref{eqn:approxdiffscaling}, \ref{eqn:TDscaling}, \ref{eqn:advscaling}). The Taylor dispersion prediction (\ref{eqn:TDscaling}) is omitted in panels (\textit{c},\textit{d}), though the predicted onset of mixing enhancement $\textit{Pe}_0$ (\ref{eqn:PeTrans}) is plotted as a vertical dashed line for all four surveys.}
    \label{fig:mixseparatesurvs}
\end{figure}

\begin{figure}
    \centering
    \includegraphics[width=1\linewidth]{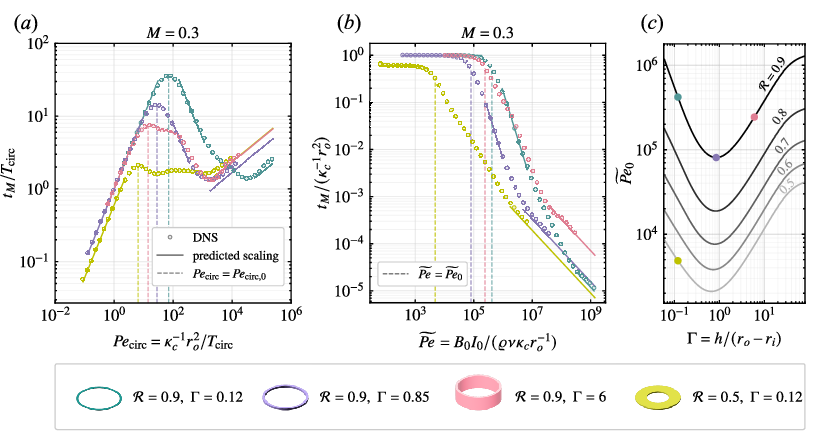}
    \caption{(\textit{a}) DNS results and asymptotic predictions for all four surveys rescaled using $T_{\text{circ}}$ to show the effects of flow morphology (rather than flow magnitude) on mixing enhancement. Predicted scalings are extrapolated to the highest value of $\textit{Pe}_{\text{circ}}$ investigated. (\textit{b}) Mixing enhancement from a practical standpoint. The data and predictions in the previous panel are rescaled such that the fluid properties and channel radius $r_o$ may be regarded as fixed while the electromagnetic forcing $B_0 I_0$ is varied. Predicted scalings are extrapolated to the highest value of $\widetilde{\textit{Pe}}$ investigated. (\textit{c}) Asymptotic predictions for the onset of mixing enhancement $\widetilde{\textit{Pe}}_0$ (using \ref{eqn:PeTrans}, \ref{eqn:PeTildeDef}) as a function of depth-to-gap-width ratio $\Gamma$ at different values of $\mathcal{R}$ (contour labels). Coloured points correspond to vertical dashed lines in the previous panel.}
    \label{fig:mixcombsurvs}
\end{figure}

We compare the asymptotic scaling predictions above to 3D DNS of  (\ref{eqn:advdiffndm}, \ref{eqn:IC}) over five orders of magnitude in $\textit{Pe}$. Three surveys in shallow, transitional, and deep magneto-Stokes regimes ($\Gamma = 0.12,0.85,6$) are explored for a channel with $\mathcal{R}=0.9$; an additional survey with $\mathcal{R}=0.5$ and $\Gamma = 0.12$ is included for comparison. All four surveys are indicated with open markers on the regime map in figure \ref{fig:regimes}(a) (colour scheme is maintained between figures \ref{fig:regimes}, \ref{fig:mixseparatesurvs}, \ref{fig:mixcombsurvs}), and the three $\mathcal{R}=0.9$ flow profiles are plotted in figure \ref{fig:regimes}(b,c). The details of the numerical method are included in Appendix \ref{app:C}.

Figure \ref{fig:mixseparatesurvs} plots the dimensionless mixing times $t_M/T$ corresponding to $M = 0.5, 0.3, 0.15$ against $\textit{Pe}$ for each DNS survey (points). The asymptotic scaling laws (\ref{eqn:approxdiffscaling}, \ref{eqn:TDscaling}, \ref{eqn:advscaling}) are plotted as solid lines for each value of $M$ (both the exponent and coefficient are predicted for these curves). The data in all four surveys follow diffusive and advection-dominated scalings. Taylor dispersion following (\ref{eqn:TDscaling}) only appears in the shallow ($\Gamma=0.12$) and transitional ($\Gamma=0.85$) surveys in the narrow channel ($\mathcal{R}=0.9$) (figure \ref{fig:mixseparatesurvs}a,b). In the shallow ($\Gamma=0.12$) survey, we find a fourth regime characterised by weak dispersion located between $\textit{Pe}^{-1}$ and $\textit{Pe}^{1/3}$ regimes, that was not observed by \cite{gleeson_modelling_2004} in $\Gamma \to \infty$ flows. For the deeper and wider channels (figure \ref{fig:mixseparatesurvs}c,d), behaviour at intermediate $\textit{Pe}$ is more complex (in the $\mathcal{R}=0.5$ survey, Taylor dispersion disappears entirely), and we do not plot (\ref{eqn:TDscaling}) for these surveys. Nonetheless, the transition from diffusive mixing to intermediate-$\textit{Pe}$ behaviour is accurately predicted by $\textit{Pe}_0$ (\ref{eqn:PeTrans}) in all four surveys (vertical dashed lines in figure \ref{fig:mixseparatesurvs}).

All four surveys are compared in figure \ref{fig:mixcombsurvs} (for $M=0.3$) to elucidate the effects of channel geometry ($\mathcal{R}$, $\Gamma$) on enhanced mixing. In figure \ref{fig:mixcombsurvs}(a), mixing times are scaled by the surface mid-gap circulation time,
\begin{equation}\label{eqn:taucircdef}
    T_{\text{circ}} = {\pi(r_i+r_o)}/{u_{\theta,\text{mid-gap}}},
\end{equation}
and the Péclet number is rescaled as
\begin{equation}\label{eqn:PeCircDef}
    \textit{Pe}_{\text{circ}} = \frac{u_{\theta,\text{mid-gap}}r_o^2}{U \pi h^2} \textit{Pe} =\frac{r_o^2/\kappa_c}{T_{\text{circ}}}.
\end{equation}
This rescaling is equivalent to considering the circulation time and outer radius to be equal for all surveys, and allowing only the height, inner radius, and diffusivity to vary; this isolates the effect of the flow's morphology (e.g., the size of boundary layers) from its magnitude.

Focusing on the $\mathcal{R}=0.9$ surveys (pink, purple, and teal points), we observe two trends in figure \ref{fig:mixcombsurvs}(a) as depth-to-gap-width ratio $\Gamma$ is decreased: (i) the onset of mixing enhancement is delayed ($\textit{Pe}_{\text{circ},0}$ increases; the dashed teal line lies to the right of the dashed pink line), and (ii) mixing efficiency in the advection-dominated regime increases ($t_M/T_{\text{circ}}$ decreases; the solid teal line lies below the solid pink line on the right side of \ref{fig:mixcombsurvs}a). The onset of mixing enhancement $\textit{Pe}_0$ is controlled by Taylor dispersion, which depends strongly on shear in the radial direction. Thus, as $\Gamma$ is decreased, the sidewall boundary layers shrink (\ref{eqn:BLscaling}), Taylor dispersion is reduced, and diffusion-dominated mixing persists to higher $\textit{Pe}$. Mixing in the advection-dominated regime occurs via diffusion across lamellar structures in the tracer concentration field, which become vertically-interleaved in flows dominated by vertical shear (large basal boundary layers). As $\Gamma$ is decreased, the basal boundary layer grows (\ref{eqn:BLbasalscaling}), the tracer concentration front is vertically-sheared into a spiraling interface, and $t_M/T_{\text{circ}}$ decreases. Finally, we observe that the onset of mixing enhancement occurs earlier in the wider channel $\mathcal{R} = 0.5$ (chartreuse points) than in the narrow channel with the same depth-to-gap-width ratio (pink points). The same trends in figure \ref{fig:mixcombsurvs}(a) are present for $M=0.15$ and $M=0.5$.

The effects of flow morphology alone would seem to advocate for the use of deeper channels, since the onset of mixing enhancement occurs sooner (holding $T_{\text{circ}}$ constant). However, achieving flow speeds in deep channels that are comparable to those in shallow channels is difficult in practice, as stronger magnetic fields or applied currents are required to counteract the drop in current density with depth. Thus, a practical representation of our mixing results requires one to combine the influence of flow morphology on mixing (figure \ref{fig:mixcombsurvs}a) with the influence of depth-to-gap-width ratio on flow magnitude (figure \ref{fig:soln}).

To this end, figure \ref{fig:mixcombsurvs}(b) plots mixing times scaled by the diffusive timescale $r_o^2/\kappa_c$ versus the Péclet number rescaled as
\begin{equation}\label{eqn:PeTildeDef}
    \widetilde{\textit{Pe}} = \frac{2 \pi (1+\mathcal{R})^2}{\mathcal{H}^3}\textit{Pe} = \frac{B_0 I_0}{\varrho \nu \kappa_c/r_o}.
\end{equation}
This rescaling allows us to observe the change in mixing time versus $\Gamma$, $\mathcal{R}$, and the electromagnetic forcing ($B_0 I_0$) for a chosen solution (constant $\varrho$, $\nu$, $\kappa_c$) and fixed outer radius ($r_o$). In the advection-dominated limit, the shallowest channels still induce the fastest mixing; in the lower right portion of figure \ref{fig:mixcombsurvs}(b), the extrapolated mixing time prediction decreases by more than five fold between the survey with $\Gamma = 0.6$, $\mathcal{R} = 0.9$ (solid pink line) and the survey with $\Gamma = 0.12$, $\mathcal{R} = 0.9$ (solid teal line). However, the enhancement of mixing is initiated with the least effort (smallest $B_0 I_0$) for the transitional flow ($\Gamma = 0.85$; purple dashed line), out of the three $\mathcal{R}=0.9$ surveys (pink, purple, and teal dashed lines).

Figure \ref{fig:mixcombsurvs}(c) plots the predicted onset value $\widetilde{\textit{Pe}}_0$ as a function of $\Gamma$ for different values of $\mathcal{R}$. This shows that the transitional flow (purple point) in fact lies at a minimum in $\widetilde{\textit{Pe}}_0$ for $\mathcal{R} = 0.9$. These optima, located between $\Gamma = 0.66$ for $\mathcal{R} = 0.5$ and $\Gamma =0.85$ for $\mathcal{R} = 0.9$, result from the competing effects of sidewall boundary layer size and flow speed as $\Gamma$ is varied and the electromagnetic parameters are kept constant. Thus, magneto-Stokes annuli with near-unity depth-to-gap-width ratios enhance mixing for the least electromagnetic forcing.

Specific mixing applications include DNA hybridisation assays (e.g., \citealt{heule_sequential_2004}), which are hindered by the extremely low chemical diffusivities typical of macromolecules in aqueous solutions \citep{gregory_magnetohydrodynamic_2016}. For example, it takes more than 5 days ($r_o^2/\kappa_c$) for 20-bp DNA fragments ($\kappa_c = 5.7\times 10^{-11}$ m$^2$/s; \citealt{lukacs_size-dependent_2000}) to diffuse through a microfabricated annulus with $r_o = 5$ mm, $r_i = 4$ mm, and $h = 425$ $\mu$m (cf. \citealt{west_structuring_2003}). In contrast, applying a modest electromagnetic forcing ($I_0 = 0.1$ A, $B_0 = 25$ mT) to this system results in advection-dominated mixing ($\widetilde{\textit{Pe}} = 2.2 \times 10^8$) with predicted mixing time $t_{M=0.3} = $ 14 s.

Actual mixing times may be further reduced by the effects of surface tension. Deflection of the free surface at inner and outer menisci may locally reduce current density, thus enlarging the sidewall boundary layers and augmenting Taylor dispersion. Though potentially useful, these complications may be avoided by placing a no-slip upper boundary at $z = 2 h$. If the total current is doubled ($I= 2 I_0$), then the equations for momentum (\ref{eqn:PDETheta}) and advection-diffusion (\ref{eqn:advdiffndm}, \ref{eqn:IC}) are unchanged, and their solutions are simply extensions of the free-surface solutions mirrored across the $z = h$ plane. Thus, the asymptotic mixing time predictions (\ref{eqn:approxdiffscaling}, \ref{eqn:TDscaling}, \ref{eqn:advscaling}) for the free-surface system  also apply to a closed system with half-height $h$ and half-current $I_0$.
\section{Discussion}\label{sec:disc}
In this study, we provide the first fully-analytical solution for a fundamental MHD flow: magneto-Stokes flow in a cylindrical-annulus. Three flow regimes (shallow-layer, transitional, and deep-layer) are distinguished based on a single geometric parameter: the depth-to-gap-width ratio $\Gamma$.

We characterise the effect of $\Gamma$ on the homogenisation of a diffusing tracer, relevant to the design of microscale mixing devices \citep{gleeson_modelling_2004, west_structuring_2003}. Mixing in infinitesimally thin layers ($\Gamma \to 0$) proceeds without the benefit of axial vorticity. In fact, we show that the shallow-layer asymptotic solution belongs to a class of MHD potential flows. These findings have already generated interest in analytical solutions for magneto-Stokes flow in other multiply-connected channels of vanishing depth \citep{mckee_exact_2024}.

In finite-depth channels, mixing efficiency depends strongly on the value of $\Gamma$. Using asymptotic reductions of the advection-diffusion equation validated with 3D DNS, we show that $\Gamma \approx 1$ annuli are optimal for initiating mixing enhancement with the least electromagnetic effort ($B_0 I_0$). If the magnitude of $B_0 I_0$ is not a constraint, then the shortest mixing times may be achieved via strongly-forced, advection-dominated mixing in shallow channels ($\Gamma \ll 1$).

The extensive characterisation of both momentum and tracer evolution provided here makes the annular magneto-Stokes system an excellent MHD reference flow. Among other applications, its promise as a calibration tool for particle tracking velocimetry (PTV) (e.g., \citealt{valenzuela-delgado_electrolyte_2018}) and particle image velocimetry (PIV) draws on the simplicity of the device and robustness of the analytical solution.

\backsection[Supplementary data]{\label{SupMat}A supplementary Python package and accompanying Jupyter notebook that implement our theoretical predictions are available at \href{https://github.com/cysdavid/magnetoStokes}{https://github.com/cysdavid/magnetoStokes}}

\backsection[Acknowledgements]{The authors thank Ernest Gomis and Andrea Chlarson for their early laboratory work on the magneto-Stokes annulus. C.S.D. thanks Ellin Zhao for her expertise in image-processing methods, which greatly improved the phase-boundary tracking code used in this study. The authors are grateful to Lingyun Ding for his advice on the centre manifold approach to modelling Taylor dispersion.}

\backsection[Funding]{This research was supported by the National Science Foundation (EAR 1620649, EAR 1853196).}

\backsection[Declaration of interests]{The authors report no conflict of interest.}

\backsection[Data availability statement]{The laboratory and DNS data that support our findings are openly available at \url{https://doi.org/10.5281/zenodo.12362602}. Additionally, our dye-tracking velocimetry program and DNS codes may be freely accessed at \url{https://github.com/cysdavid/magnetoStokes}. A Mathematica notebook that reproduces our analytical solution is also included in this Github repository.}

\backsection[Author ORCIDs]{C.S. David, https://orcid.org/0009-0006-9774-1471; E.W. Hester, https://orcid.org/0000-0003-1651-9141; Y. Xu, https://orcid.org/0000-0001-9123-124X; J.M. Aurnou, https://orcid.org/0000-0002-8642-2962}

\backsection[Author contributions]{C.S.D. developed the model, found the solution presented here, and derived the asymptotic predictions for mixing time. E.W.H. wrote the DNS codes for both the axisymmetric 2D standard cases and the 3D mixing cases; C.S.D ran these codes. C.S.D. and Y.X. carried out laboratory experiments, with Y.X. providing key insights into flow visualisation. J.M.A. conceived the idea for this study and guided its development. C.S.D. prepared all drafts of this manuscript, and all authors contributed to the final version.}

\appendix

\section{Full axisymmetric equations and solution}\label{appA}
The 3D axisymmetric equations under the nondimensionalisation in \S\ref{sec:theory:gov} are
\begin{subequations}\label{eqn:3Dgov}
\begin{align}
    \Rey &\left\{\partial_{\tau}\upsilon_\rho+(\mathcal{R} +1) \left[\frac{1}{\mathcal{H}}(\boldsymbol{\upsilon}_\bot\bcdot{\bnabla}_\bot)\upsilon_\rho-\frac{\upsilon_{\theta }^2}{\rho}+\partial_{\rho}\Pi\right]\right\}={\nabla}^{2}_{\bot} \upsilon_\rho - \mathcal{H}^2\frac{\upsilon_\rho}{\rho^2},\\
    \Rey &\left\{\partial_{\tau}\upsilon_{\theta }+(\mathcal{R} +1) \left[\frac{1}{\mathcal{H}}(\boldsymbol{\upsilon}_\bot\bcdot{\bnabla}_\bot)\upsilon_\theta+\frac{\upsilon_\rho \upsilon_{\theta }}{\rho}\right]\right\}={\nabla}^{2}_{\bot} \upsilon_\theta - \mathcal{H}^2\frac{\upsilon_\theta}{\rho^2}
    +\frac{\mathcal{R} +1}{\rho}\Upsilon(\tau),\\
    \Rey &\left\{\partial_{\tau}\upsilon_\zeta+({\mathcal{R} +1})\left[\frac{1}{\mathcal{H}}(\boldsymbol{\upsilon}_\bot\bcdot{\bnabla}_\bot)\upsilon_\zeta +\frac{1}{\mathcal{H}^2}{\partial_{\zeta}\Pi}\right]\right\}={\nabla}^{2}_{\bot} \upsilon_\zeta,
\end{align}
\end{subequations}

\begin{equation}\label{eqn:3Dcont}
    \partial_{\rho}(\rho\upsilon_\rho) + {\rho}\partial_{\zeta}\upsilon_\zeta =0,
\end{equation}
where we define $\boldsymbol{\upsilon}_\bot = \upsilon_\rho \boldsymbol{e_r} + \mathcal{H}\upsilon_\zeta \boldsymbol{e_z}$, ${\bnabla}_\bot (\cdot)= \left[\boldsymbol{e_r}\mathcal{H}\partial_{\rho} + \boldsymbol{e_z}\partial_{\zeta}\right](\cdot)$, and ${\nabla}^{2}_{\bot}(\cdot) = \left[\mathcal{H}^2 {\rho}^{-1} \partial_{\rho}{\rho}\partial_{\rho} + \partial_{\zeta}^2\right](\cdot)$. For vanishing meridional flow $\boldsymbol{\upsilon}_\bot$, (\ref{eqn:3Dgov}) reduces to (\ref{eqn:PDETheta}).

In \S\ref{sec:theory:gov}, we provide an expression for $\upsilon_\theta$ found by approximating the full solution to (\ref{eqn:PDETheta}):
\begin{eqnarray}\label{eqn:fullIVPsoln}
\upsilon_\theta(\rho,\zeta,\tau) & =&  \sum_{n=1}^{\infty} \Biggr\{\frac{2(\mathcal{R}+1)}{k_n^2 \mathcal{H}} \left[ \frac{\mathcal{H}}{k_n \rho}-A_n \rmI_1\left( \frac{k_n}{ \mathcal{H}}\rho\right) - B_n \rmK_1\left( \frac{k_n}{ \mathcal{H}}\rho\right) \right] \\
    &&
+\sum_{m=1}^{\infty}C_{mn}\left(\frac{\rmJ_ 1\left(\mu_m  \rho\right)}{\rmJ_1(\mu_m)}-	\frac{\rmY_1\left(\mu_m  \rho		\right)}{\rmY_ 1(\mu_m )}\right) 
	\exp\left({-\frac{\mathcal{H} ^2 \mu_m ^2+k_n^2}{\Rey}\tau }\right)\Biggr\}\sin \left(k_n\zeta\right)\nonumber,
\end{eqnarray}
where $\rmJ_1$ and $\rmY_1$ are first-order Bessel functions of first and second kind and each eigenvalue $\mu_m$ is the $m$-th smallest positive root of
\begin{equation}\label{eqn:solvCond}
\rmY_1(\mu ) \rmJ_1(\mu  \mathcal{R} )-\rmJ_1(\mu ) \rmY_1(\mu  \mathcal{R})=0.
\end{equation}

Setting $\mathcal{H} ^2 \mu_m ^2+k_n^2 \approx k_1^2$ in the exponential term above reduces (\ref{eqn:fullIVPsoln}) to the approximate solution (\ref{eqn:combSoln}). The roots of (\ref{eqn:solvCond}) and analytical expressions for coefficients $C_{m n}$ are provided in a Mathematica notebook available at \href{https://github.com/cysdavid/magnetoStokes}{https://github.com/cysdavid/magnetoStokes}.

\section{Derivation of mixing time predictions}\label{app:B}
Derivation of both the diffusion-dominated (\ref{eqn:approxdiffscaling}) and advection-dominated (\ref{eqn:advscaling}) scaling laws is facilitated by transforming (\ref{eqn:advdiffndm}) into the Lagrangian reference frame ($\rho$,$\tilde{\theta}$,$\zeta$) according to $\theta = \tilde{\theta} + (1+\mathcal{R})\omega(\rho,\zeta) \tau$. We assume that the mixing time scales with some power $\alpha$ of the Péclet number, and we let $\tau = \textit{Pe}^\alpha \tilde{\tau}$ so that (\ref{eqn:advdiffndm}) becomes
\begin{equation}
\begin{split}
    \partial_{\tilde{\tau}} c = \textit{Pe}^{\alpha-1}\left(\frac{\mathcal{H}^2} {\rho^{2}}\partial_{\tilde{\theta}}^2 + {\nabla}^{2}_{\bot} \right)c \;-\;& \textit{Pe}^{2\alpha-1}(1+\mathcal{R})\tilde{\tau}\left[{\nabla}^{2}_{\bot}\omega + 2({\bnabla}_\bot\omega \cdot {\bnabla}_\bot)\right]\partial_{\tilde{\theta}}c\\ 
    +\;& \textit{Pe}^{3\alpha-1}(1+\mathcal{R})^2 \tilde{\tau}^2 \left( {\bnabla}_\bot\omega \cdot {\bnabla}_\bot\omega\right) \partial_{\tilde{\theta}}^2 c.
\end{split}
\end{equation}
    
In the limit $\textit{Pe} \to 0$, a dominant balance exists if $\alpha = 1$. This results in the classical diffusion equation, which yields (\ref{eqn:approxdiffscaling}) upon solution. On the other hand, two-term dominant balance in the limit $\textit{Pe} \to \infty$ requires that $\alpha = 1/3$, yielding a diffusion-like equation:
\begin{equation}\label{eqn:hiPeadvdiff}
    \partial_{\tilde{\tau}} c = (1+\mathcal{R})^2 \tilde{\tau}^2 \left( {\bnabla}_\bot\omega \cdot {\bnabla}_\bot\omega\right) \partial_{\tilde{\theta}}^2 c,
\end{equation}
whose solution,
\begin{equation}\label{eqn:advasympsoln}
    c(\rho,\tilde{\theta},\zeta,\tilde{\tau}) = \overline{c} + \sum_{m=1}^\infty C_m \sqrt{2} \cos{m \tilde{\theta}} \; e^{-\frac{1}{3}(1+\mathcal{R})^2\left( {\bnabla}_\bot\omega \cdot {\bnabla}_\bot\omega\right)m^2 \tilde{\tau}^3},
\end{equation}
depends on $\rho$ and $\zeta$ parametrically through the angular velocity $\omega$. The advection-dominated scaling law (\ref{eqn:advscaling}) follows from retaining only the fundamental ($m=1$) mode in (\ref{eqn:advasympsoln}).

The Taylor dispersion scaling prediction is derived using a separate procedure. We transform (\ref{eqn:advdiffndm}) according to  $\theta = \vartheta + (1+\mathcal{R})\langle {\omega} \rangle\tau$ into the frame ($\rho$,$\vartheta$,$\zeta$) rotating with the average angular velocity $\langle \omega \rangle$. In this frame, we assume that mixing smooths out the cross-sectional averaged tracer concentration field $\langle c \rangle$ such that 
\begin{equation}
    \langle c \rangle \gg \partial_\vartheta \langle c \rangle \gg \partial_\vartheta^2 \langle c \rangle \gg \partial_\vartheta^3 \langle c \rangle \gg...
\end{equation}
as $\tau \to \infty$. Given this ordering, we may effect a centre-manifold reduction \citep{mercer_complete_1994,roberts_low-dimensional_1996,ding_determinism_2022} of the 3D advection-diffusion equation by adopting an asymptotic expansion for $c$ similar to that of \cite{ding_diffusion-driven_2023}:
\begin{equation}\label{eqn:centmanansatz}
    c = \langle c \rangle + \textit{Pe} \;a_1(\rho,\zeta)\frac{1}{\rho_0}\partial_\vartheta \langle c \rangle + \textit{Pe}\; a_2(\rho,\zeta)\frac{1}{\rho_0^2}\partial_\vartheta^2 \langle c \rangle + O(\partial_\vartheta^3 \langle c \rangle),
\end{equation}
where $\rho_0 = (1+\mathcal{R})/2$ is the mid-gap radius. The functions $a_1, a_2$ vanish upon cross-sectional averaging ($\langle a_1 \rangle = \langle a_2 \rangle=  0$) and must satisfy the same no-flux boundary conditions as does $c$ (such that $\langle {\nabla}^{2}_{\bot}a_1 \rangle = \langle {\nabla}^{2}_{\bot}a_2 \rangle=  0$).

Substituting (\ref{eqn:centmanansatz}) into the transformed advection-diffusion equation yields
\begin{multline}\label{eqn:advdiffasympexpand_2D}
    \partial_\tau\left(\langle c \rangle + \textit{Pe}\;\frac{a_1}{\rho_0}\partial_\vartheta \langle c \rangle + \textit{Pe}\;\frac{a_2}{\rho_0^2}\partial_\vartheta^2 \langle c \rangle\right)+(1+\mathcal{R})(\omega-\langle {\omega} \rangle)\left(\partial_\vartheta \langle c \rangle+\textit{Pe}\;\frac{a_1}{\rho_0}\partial_\vartheta^2 \langle c \rangle\right) \\= \frac{\partial_\vartheta \langle c \rangle}{\rho_0}{\nabla}^{2}_{\bot} a_1 + \frac{\partial_\vartheta^2 \langle c \rangle}{\rho_0^2}{\nabla}^{2}_{\bot} a_2 + O(\partial_\vartheta^3 \langle c \rangle),
\end{multline}
which, after cross-sectional averaging $\langle \cdot \rangle$ and neglecting $O(\partial_\vartheta^3 \langle c \rangle) $ terms, leaves a 1D diffusion equation for $\langle c \rangle(\vartheta,\tau)$:
\begin{equation}\label{eqn:taylordispeqn}
    \partial_\tau \langle c \rangle = \mathcal{C}_D \textit{Pe}\frac{1}{\rho_0^2}\partial_\vartheta^2\langle c \rangle,
\end{equation}
where $\mathcal{C}_D$ depends on $\omega$ and $a_1$ through (\ref{eqn:TaylorDispCoeff}). To determine $a_1$ and obtain closure, we first observe from (\ref{eqn:taylordispeqn}) that $\partial_\tau \langle c \rangle =O( \partial_\vartheta^2 \langle c \rangle)$. Then, collecting $O(\partial_\vartheta \langle c \rangle)$ terms in (\ref{eqn:advdiffasympexpand_2D}) yields the Poisson equation (\ref{eqn:a1_2D}), which may be solved for $a_1$. Finally, the advection-dominated scaling law (\ref{eqn:advscaling}) follows from solving (\ref{eqn:taylordispeqn}) and retaining only the $m=1$ mode.

\section{Numerical method for 3D mixing DNS}\label{app:C}
The 3D simulations of the advection-diffusion equation (\ref{eqn:advdiffndm}) were performed using the pseudospectral code Dedalus. The flow field $u_\theta$ was computed using the first $\mathcal{N}$ vertical modes of the steady analytical flow solution (\ref{eqn:steadySoln}), truncated such that the $\mathcal{N}+1$st mode changes the value of $u_{\theta,\text{mid-gap}}$ by $<$0.5\%. The tracer concentration $c$ and velocity $u_\theta$ fields were discretised using Chebyshev modes in the $\rho$, $\zeta$ directions and real Fourier modes in the $\theta$ direction. The azimuthally discontinuous initial condition for $c$ (\ref{eqn:IC}) was approximated by a smooth bump function to avoid Gibbs oscillations:
\begin{equation}
c_0(\rho,\theta) =
\begin{cases}
 0, & \left(\frac{\pi }{2} + \frac{\Delta \theta}{2 \rho}\right)^2\leq (\theta -\pi )^2 \\
 S\left(\theta -\pi; \frac{\pi }{2}-\frac{\Delta \theta}{2 \rho},\frac{\Delta
   \theta}{\rho}\right), & \left(\frac{\pi }{2}-\frac{\Delta \theta}{2 \rho}\right)^2<(\theta -\pi
   )^2<\left(\frac{\pi }{2}+\frac{\Delta \theta}{2 \rho}\right)^2 \\
 1, & (\theta -\pi )^2\leq \left(\frac{\pi }{2}-\frac{\Delta \theta}{2 \rho}\right)^2
\end{cases},
\end{equation}
where $S$ is a transition function constructed following \cite{tu_introduction_2011},
\begin{equation}
    S(x;w,\Delta w) = \left\{{1+\exp \left[\frac{\Delta w (\Delta w+2 w) \left(w^2+(\Delta w+w)^2-2 x^2\right)}{\left(x^2-w^2\right) \left(x^2-(\Delta w+w)^2\right)}\right]}\right\}^{-1},
\end{equation}
and the ramp width $\Delta \theta$ is set to $\Delta \theta = 2\pi (12/256)$ across all cases.

For efficiency reasons, we used different timestepping schemes in cases dominated by diffusion versus those with stronger advection (see \citealt{ascher_implicit-explicit_1997}). The empirically determined cutoff value $\textit{Pe}^*$ between these two groups roughly coincides with $\textit{Pe}_0$, the predicted transition between diffusion-dominated and Taylor dispersion mixing regimes. The strongly diffusive cases ($\textit{Pe} < \textit{Pe}^*$) used the SBDF2 scheme \citep{wang_variable_2008}, while the remainder ($\textit{Pe} \ge \textit{Pe}^*$) used the four-stage, third-order implicit-explicit Runge-Kutta (RK443) scheme (\citealt{ascher_implicit-explicit_1997}, \S2.8) and twice the spatial resolution $(N_\rho, N_\theta,N_\zeta)$ of the diffusive cases. These values, in addition to  $\textit{Pe}^*$, $\mathcal{N}$, and the timestep $\Delta \tau$ for each survey are collected in table \ref{tab:3DDNS}. The code used for these simulations is available online (\url{https://github.com/cysdavid/magnetoStokes}).
\begin{table}
\begin{center}
\addtolength{\tabcolsep}{2pt} 
\begin{tabular}{cclcccc}
$\Gamma$ & $\mathcal{R}$ &  $\mathcal{N}$&$\textit{Pe}^*$ &$(N_\rho, N_\theta,N_\zeta) $ for $\textit{Pe} \ge \textit{Pe}^*$& $\Delta \tau$ ($\textit{Pe} \ge \textit{Pe}^*$) & $\Delta \tau$ ($\textit{Pe} < \textit{Pe}^*$) \\ \hline
0.12     & 0.5           &  3&0.1             &(128, 256, 64)& 0.0039                                          & 0.0033                                        \\
0.12     & 0.9           &  3&0.14            &(128, 256, 64)& 0.0057                                          & 0.0057                                        \\
0.85     & 0.9           &  6&10              &(128, 256, 64)& 0.02                                            & 0.02                                          \\
6        & 0.9           &  46&$10^4$          &(64, 256, 128)& 0.84                                            & 0.84                                         
\end{tabular}
\caption{Numerical parameters used for the 3D DNS discussed in \S\ref{sec:apps:mixing}, including the threshold Péclet number $\textit{Pe}^*$ below which SBDF2 timestepping is used instead of RK443. Spatial resolution ($N_\rho$, $N_\theta$, $N_\zeta$) for cases with $\textit{Pe} < \textit{Pe}^*$ is half that of the values shown here for each survey.}
\label{tab:3DDNS}
\end{center}
\end{table}
\newpage
\bibliographystyle{jfm}
\expandafter\ifx\csname natexlab\endcsname\relax
\def\natexlab#1{#1}\fi
\expandafter\ifx\csname selectlanguage\endcsname\relax
\def\selectlanguage#1{\relax}\fi
\bibliography{references}

\end{document}